\documentclass[aps,prx,twocolumn,superscriptaddress,longbibliography,floatfix]{revtex4-1}

\usepackage{amsmath}
\usepackage{graphicx}
\usepackage[colorlinks,linkcolor=blue,citecolor=blue,urlcolor=blue]{hyperref}
\usepackage[dvipsnames]{xcolor}
\usepackage{stix}
\usepackage[normalem]{ulem}
\begin{document}

\newcommand{\dd}{\mathrm{d}}
\renewcommand{\vec}[1]{\mathbf{#1}}
\newcommand{\gvec}[1]{\boldsymbol{#1}}
\newcommand{\en}{\varepsilon}
\newcommand{\hc}{\hat{c}}{}
\newcommand{\hcd}{\hat{c}^\dagger}
\newcommand{\hd}{\hat{d}}
\newcommand{\hdd}{\hat{d}^\dagger}

\mathchardef\mhyphen="2A
\newcommand{\dx}[1]{\!\mathop{\mathrm{d}}\!#1\,}
\let\OldRe\Req
\renewcommand{\Re}{\text{Re}\,}
\let\OldIm\Im
\renewcommand{\Im}{\text{Im}\,}
\newcommand{\iu}{\mathrm{i}}

\newcommand{\intd}[1]{\int\!\!\dd #1\,}
\newcommand{\iintd}[2]{\int\!\!\dd #1\!\!\int\!\!\dd #2\,}

\newcommand{\added}[1]{\textcolor{WildStrawberry}{#1}}
\newcommand{\tocheck}[1]{\textcolor{red}{#1}}

\newcommand{\qw}[3]{\sout{\textcolor{red}{#1}} \textcolor{blue}{#2} \textcolor{green}{[#3]}}


\author{Michael Sch\"uler}
\affiliation{Stanford Institude for Materials and Energy Sciences (SIMES),
SLAC National Accelerator Laboratory, Menlo Park, CA 94025, USA}
\author{Umberto De Giovannini}
\affiliation{Nano-Bio Spectroscopy Group,  Departamento de Fisica de Materiales, 
Universidad del Pa\'i{}s Vasco UPV/EHU- 20018 San Sebasti\'a{}n, Spain}
\affiliation{Max Planck Institute for the Structure and Dynamics of
  Matter, Luruper Chaussee 149, 22761 Hamburg, Germany}
\author{Hannes H\"ubener}
\affiliation{Max Planck Institute for the Structure and Dynamics of
  Matter, Luruper Chaussee 149, 22761 Hamburg, Germany}
\author{Angel Rubio}
\affiliation{Max Planck Institute for the Structure and Dynamics of
  Matter, Luruper Chaussee 149, 22761 Hamburg, Germany}
\affiliation{Center for Computational Quantum Physics (CCQ), 
  The Flatiron Institute, 162 Fifth avenue, New York NY 10010}
\affiliation{Nano-Bio Spectroscopy Group,  Departamento de Fisica de Materiales, 
Universidad del Pa\'i{}s Vasco UPV/EHU-20018 San Sebasti\'a{}n, Spain}
\author{Michael A. Sentef}
\affiliation{Max Planck Institute for the Structure and Dynamics of
  Matter, Luruper Chaussee 149, 22761 Hamburg, Germany}
\author{Thomas P. Devereaux}
\affiliation{Stanford Institude for Materials and Energy Sciences (SIMES),
SLAC National Accelerator Laboratory, Menlo Park, CA 94025, USA}
\affiliation{Department of Materials Science and Engineering, Stanford University, Stanford, California 94305, USA}
\author{Philipp Werner}
\affiliation{Department of Physics, University of Fribourg, 1700 Fribourg, Switzerland}

\title{How Circular Dichroism in time- and angle-resolved photoemission can be used to spectroscopically detect transient topological states in graphene}

\begin{abstract}
  Pumping graphene with circularly polarized light is the archetype of light-tailoring topological bands. Realizing the induced Floquet-Chern insulator state and tracing clear experimental manifestions has been a challenge, and it has become clear that scattering effects play a crucial role. We tackle this gap between theory and experiment by employing microscopic quantum kinetic calculations including realistic electron-electron and electron-phonon scattering. Our theory provides a direct link to 
  the build-up of the Floquet-Chern insulator state in light-driven graphene and its detection  
  in time- and angle-resolved 
  photoemission spectroscopy (ARPES). This allows us to study the stability of the Floquet features due to dephasing and thermalization effects. We also discuss the ultrafast Hall response in the laser-heated state. Furthermore, the induced pseudospin texture and the associated Berry curvature gives rise to momentum-dependent orbital magnetization, which is reflected in circular dichroism in ARPES (CD-ARPES). Combining our nonequilibrium calculations with an accurate one-step theory of photoemission allows us to establish a direct link between the build-up of the topological state and the dichroic pump-probe photoemission signal. The characteristic features in CD-ARPES are further corroborated to be stable against heating and dephasing effects. Thus, tracing circular dichroism in time-resolve photoemission provides new insights into transient topological properties.
\end{abstract}
\pacs{abstract}
\maketitle

\section{Introduction}
Topological properties play an important role in the study of fundamental phenomena in condensed matter systems. 
In periodic systems, the notion of quantum-geometric properties like the Berry curvature and their implications on the macroscopic scale has become a central concept. The most prominent examples are topological insulators (TIs) and superconductors~\cite{hasan_colloquium:_2010,qi_topological_2008} with their protected surface or edge states. Realizing topological insulators with integer quantum anomalous Hall effect (QAHE) has proven to be a challenge. In this regard, the remarkable progress in creating two-dimensional (2D) materials and heterostructures thereof has opened new perspectives~\cite{kou_two-dimensional_2017,marrazzo_prediction_2018,tang_quantum_2017,muechler_topological_2016}. In 2D materials, the topology typically arises due to the Kane-Mele mechanism~\cite{kane_quantum_2005,kane_$z_2$_2005}: a gap opens at Dirac cones due to spin-orbit coupling, giving rise to band inversion and thus a topologically nontrival state. Monolayer graphene is a paradigmatic example, and many attempts have been made to turn graphene into a TI~\cite{weeks_engineering_2011,castro_neto_impurity-induced_2009,abdelouahed_spin-split_2010}. 

The possibility of opening a gap \emph{dynamically} by pumping graphene with circularly polarized light has first been proposed in Ref.~\cite{oka_photovoltaic_2009}. In a Floquet picture, the periodic electric field renormalizes the band structure by virtual photon emission and absorption processes. By tailoring the pump frequency and strength, a Chern-insulating phase can be induced (Floquet-Chern insulator) \cite{lindner_floquet_2011}, which shows features of a QAHE~\cite{mikami_brillouin-wigner_2016,dehghani_out--equilibrium_2015,dehghani_optical_2015}. The concept of topological states engineered by periodic driving fields has been extended to experiments on ultracold atoms~\cite{jotzu_experimental_2014,dalessio_dynamical_2015}, coherent excitations of the lattice degrees of freedom~\cite{hubener_phonon_2018}, more general classes of 2D materials~\cite{claassen_all-optical_2016}, and different topological states such as Dirac and Weyl semimetals~\cite{hubener_creating_2017}. 

\begin{figure}[b]
    \centering
    \includegraphics{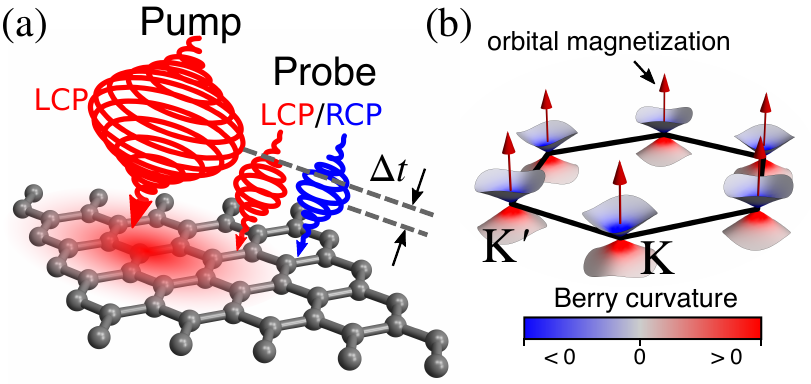}
    \caption{(a) Sketch of a pump-probe trARPES setup: a left-handed circularly polarized (LCP) pulse transiently dresses the electronic structure, which is probed by a right-handed circularly polarized (RCP), or LCP, short probe pulse. Delaying the probe pulse by $\Delta t$ with respect to the pump provides access to real-time dynamics. (b) The induced Floquet-Chern insulator is characterized by nonzero Berry curvature in the lower effective band with the same sign at the two inequivalent Dirac points. The resulting net orbital magnetization gives rise to circular dichroism in photoemission spectroscopy.}
    \label{fig:setup}
\end{figure}

To trace the pump-induced transient changes of the structure, time- and angle-resolved photoelectron spectroscopy (trARPES) has been established as a state-of-art tool, which is well suited to capture Floquet physics~\cite{de_giovannini_monitoring_2016}. Still, observing light-induced topological phases in experiments is a challenge.
So far, the distinct features of a Floquet state -- the gap opening and Floquet sidebands (replicas of the band structure associated with the absorbtion or emission of photons) -- have only been reported for Bi$_2$Se$_3$~\cite{wang_observation_2013,mahmood_selective_2016}, although related effects like the dynamical Stark effect~\cite{reutzel_coherent_2019} or photo-dressed effective band structures~\cite{vu_light-induced_2004} have been observed for different systems.
As an alternative technique for detecting the induced topological state, time-resolved transport experiments on graphene show a pump-induced Hall response~\cite{mciver_light-induced_2020}. However, in these experiments, a pump-induced population imbalance also plays a role~\cite{sato_microscopic_2019-1,sato_light-induced_2019-1}, and disentangling such effects from those of the induced Berry curvature is a nontrivial task. 
Furthermore, from these studies it becomes evident that scattering effects are crucial in graphene. In particular, the associated heating and dephasing effects compete with the coherence required for a Floquet state~\cite{kandelaki_many-body_2018,sato_floquet_2020}, although it is not clear yet which scattering mechanism is most important. 

We address this challenge in the present work by considering the pump-induced dynamics in graphene including both electron-electron (e--e) and electron-phonon (e--ph) scattering. While e--e scattering determines the initial stages of thermalization~\cite{gierz_tracking_2015} and is thus essential for the theoretical description, e--ph coupling is typically responsible for the relaxation of excited states back to equilibrium on a time scale of several hundred femtoseconds to picoseconds~\cite{gierz_electronic-structural_2016,na_direct_2019-1,caruso_photoemission_2020}. 
In a pumped system far from equilibrium, ultrafast e--ph scattering furthermore plays an important role for population dynamics~\cite{allen_theory_1987,haug_quantum_1990,gadermaier_electron-phonon_2010,sentef_examining_2013,yang_inequivalence_2015,rameau_energy_2016,kabanov_electron-electron_2020}.
We focus explicitly on the experimentally relevant regime of weak to moderate pump field strength and map out the stability of the Floquet physics. Our many-body treatment is combined with a full treatment of the photoemission process, thus providing a \emph{predictive} link to trARPES. 

Besides mapping out the momentum-dependent band structure, angle-resolved photoemission spectroscopy (ARPES) can provide insights into the quantum properties of the initial state by exploiting the light polarization. For instance, the electron chirality and the pseudospin properties can give rise to distinct circular dichroism in graphene~\cite{liu_visualizing_2011,gierz_graphene_2012} or TI surface states~\cite{wang_circular_2013,park_chiral_2012}. More generally, momentum-dependent circular dichroism allows to trace \emph{orbital} angular momentum, which is intimately linked to the Berry curvature~\cite{razzoli_selective_2017,cho_experimental_2018}. The latter was rigorously mapped out for paradigmatic 2D systems -- including graphene -- in Ref.~\cite{schuler_local_2020}. Measuring dichroism in trARPES will provide unprecedented insights into pump-induced topological properties~\cite{volckaert_momentum-resolved_2019}. 

It is the main focus of the present work to clarify this connection. Based on our predictive theory for trARPES, which is combined with an accurate one-step calculation of the photoemission matrix elements, we map out the \emph{induced} circular dichroism in laser-driven monolayer graphene. A comprehensive analysis of the Floquet state and its stability against interaction effects reveals that both e--e and e--ph scattering play an important role. Despite the thus reduced coherence, the circular dichroism is found to be robust even in the presence of strong dissipation, where other signatures of a Floquet state -- that is, opening of a gap and side bands -- are strongly suppressed. 
Hence, the circular dichroism is a hallmark manifestation of the induced Floquet topological state, which provides conclusive insights where other methods struggle. 

\section{Setup, model and methods\label{sec:methods}}

The dynamics in graphene is modeled by the Hamiltonian
\begin{align}
    \label{eq:ham_gen}
    \hat{H}(t) = \hat{H}_0(t) + \hat{H}_\mathrm{e-e} + \hat{H}_\mathrm{e-ph} \ ,
\end{align}
where $\hat{H}_0(t)$ describes the free electronic structure including the light-matter interaction. 
We consider the next-nearest-neighbor
tight-binding (TB) model, defined by 
\begin{align}
	\label{eq:ham0_td}
  \hat{H}_0(t) = \sum_{\vec{k}}\sum_{j,j^\prime,\sigma} h_{j j^\prime}(\vec{k}-\vec{A}_\mathrm{p}(t)) \hat{c}^\dagger_{\vec{k}j \sigma} \hat{c}_{\vec{k} j^\prime \sigma} \ .
\end{align}
Here, $h_{j j^\prime}(\vec{k})$ is the TB Hamiltonian in the subspace of $p_z$ orbitals, while the pump pulse (vector potential $\vec{A}_\mathrm{p}(t)$) is incorporated via the Peierls substitution. Details are presented in Appendix~\ref{sec:tbdetails}.

The second term in Eq.~\eqref{eq:ham_gen} describes the electronic-electronic (e--e) interaction. Scattering effects are taken into account at the level of an optimized Hubbard model ($U=1.6 J$ in units of the hopping amplitude $J$), which has been shown to accurately capture the electronic structure close to equilibrium~\cite{schuler_optimal_2013}. We also include electron-phonon (e--ph) scattering by the last term in the Hamiltonian~\eqref{eq:ham_gen}, taking the full dispersion of the transverse and longitudinal acoustic and optical modes into account. The matrix-elements for the e--ph coupling can be obtained from the TB model~\cite{sohier_phonon-limited_2014,sohier_density-functional_2015-1}.

The dynamics of the interacting system is treated efficiently within the time-dependent nonequilibrium Green's functions (td-NEGF) approach~\cite{stefanucci_nonequilibrium_2013}. Due to the relatively weak correlation effects, the second-order treatment with respect to the e--e and e--ph interaction provides an accurate description. Furthermore, we employ the generalized Kadanoff-Baym ansatz (GKBA)~\cite{lipavsky_generalized_1986}, which reduces the otherwise enormous computational demands significantly, while retaining a good accuracy, as demonstrated in recent benchmark calculations~\cite{schlunzen_nonequilibrium_2017,schuler_quench_2019,murakami_ultrafast_2019}. Spectral properties are improved by spectral corrections to the GKBA~\cite{latini_charge_2014}. All details on the methods can be found in Appendix~\ref{sec:numdetails}.
	
\subsection{Time-resolved photoemission\label{subsec:trarpes_theory}}

The td-NEGF approach provides a direct link to trARPES~\cite{freericks_theoretical_2009,sentef_examining_2013} by 
\begin{align}
  \label{eq:tdarpes}
  I(\vec{k},\en_f,\Delta t) = &\mathrm{Im}\sum_{j j^\prime} \int^\infty_0\!d t\! \int^{t}_0\!d t^\prime\, s(t)s(t^\prime) M^*_{j}(\vec{k},p_\perp) \nonumber \\ &\quad  \times G^<_{jj^\prime}(\vec{k};t,t^\prime)  M_{j^\prime}(\vec{k},p_\perp) e^{-\iu \Phi(t,t^\prime)} \ ,
\end{align}
where $\Phi(t,t^\prime) = \int^t_{t^\prime}\! d \bar{t}\, [\omega_\mathrm{pr}-\en_\vec{p}(\bar t)]$. Equation~\eqref{eq:tdarpes} represents a time-dependent generalization of the one-step photoemission intensity~\cite{schattke_solid-state_2008}: the transient electronic structure of the initial states is captured by the lesser Green's function $G^<_{jj^\prime}(\vec{k};t,t^\prime)$ (obtained from the td-NEGF framework), while the coupling to the final states is determined by the matrix elements $M_j(\vec{k},p_\perp)$. We compute $M_j(\vec{k},p_\perp)$ by combining the TB model with a one-step theory of photoemission. Benchmarks against state-of-the-art calculations based on the time-dependent density functional theory~\cite{scrinzi_t_2012,DeGiovannini:2016bb,de_giovannini_monitoring_2016} ensure the predicitive power of our approach (see Appendix~\ref{sec: one-step}).
The photoelectron momentum $\vec{p}=(\vec{k},p_\perp)$ determines the energy in the absence of the pump pulse by $\en_f = \vec{p}^2/2$, while $\en_\vec{p}(t) = (\vec{p}-\vec{A}_\mathrm{p}(t))^2/2$ during the pump; the time-dependent phase factor $\Phi(t,t^\prime)$ takes the streaking of the continuum (laser-assisted photoemission, LAPE~\cite{miaja-avila_laser-assisted_2006}) into account.
The probe pulse is characterized by the central frequency $\omega_\mathrm{pr}$ and the pulse envelope $s(t)$. We denote the delay between the pump and probe pulse by $\Delta t$ (see Fig.~\ref{fig:setup}(a)).

The pump photon energy is taken as $\omega_\mathrm{p} = 1.5$~eV, while the peak field strength is chosen between $E_0=1\times 10^{-3}$ and $E_0 = 4\times 10^{-3}$ atomic units (a.\,u.) ($E_0 \simeq  0.05$~V/\AA\, to $E_0\simeq 0.2$~V/\AA), corresponding to $I_0 = 3.5\times 10^{10}$~W cm$^{-2}$ to $I_0 = 5.6\times 10^{11}$~W cm$^{-2}$ peak intensity. The largest field strength is slightly above that of experimentally achievable pulses, but reveals the physics particlularly clearly. All findings are generic and present also for weaker fields.
The pump pulse is assumed to be left-handed circularly polarized (LCP) (see Fig.~\ref{fig:setup}(a)), while we choose the envelope to contain 20 optical cycles ($T_\mathrm{p}=55$~fs duration) unless stated otherwise. 
The probe pulse is assumed to have the envelope $s(t) = \sin^2(\pi (t-\Delta t)/T_\mathrm{pr})$ with a pulse length of $T_\mathrm{pr}=26$~fs. Its polarization is assumed to be either right-handed circularly polarized (RCP) or LCP. We compute the corresponding trARPES intensity $I_\mathrm{LCP/RCP}(\vec{k},\en_f,\Delta t)$ according to Eq.~\eqref{eq:tdarpes}, thus yielding the dichroic  $I_\mathrm{CD}(\vec{k},\en_f,\Delta t)=I_\mathrm{LCP}(\vec{k},\en_f,\Delta t) - I_\mathrm{RCP}(\vec{k},\en_f,\Delta t)$ and unpolarized signal $I_\mathrm{tot}(\vec{k},\en_f,\Delta t)=I_\mathrm{LCP}(\vec{k},\en_f,\Delta t) + I_\mathrm{RCP}(\vec{k},\en_f,\Delta t)$. For the photon energy of the probe pulse we choose the value $\hbar \omega_\mathrm{pr}=52$~eV, which is sufficient to detect photoelectrons from the Diract points. Furthermore, scattering effects from the lattice have been found to be minimal at this energy, such that the intrinsic dichroism dominates~\cite{gierz_graphene_2012}.

The full time-dependent treatment based on the trARPES expression~\eqref{eq:tdarpes} is complemented by a Floquet theory in the steady-state regime, where we assume that each lattice site is coupled to a thermalizing bath. 
This provides a generic dissipation channel, which allows to investigate dephasing and dissipation effects beyond e--e and e--ph scattering. Full details are presented in Appendix~\ref{sec:floquetarpes}.

\subsection{Induced pseudospin texture and topological properties\label{subsec:heff_pseudo}}

\begin{figure}[t]
  \centering
  \includegraphics[width=\columnwidth]{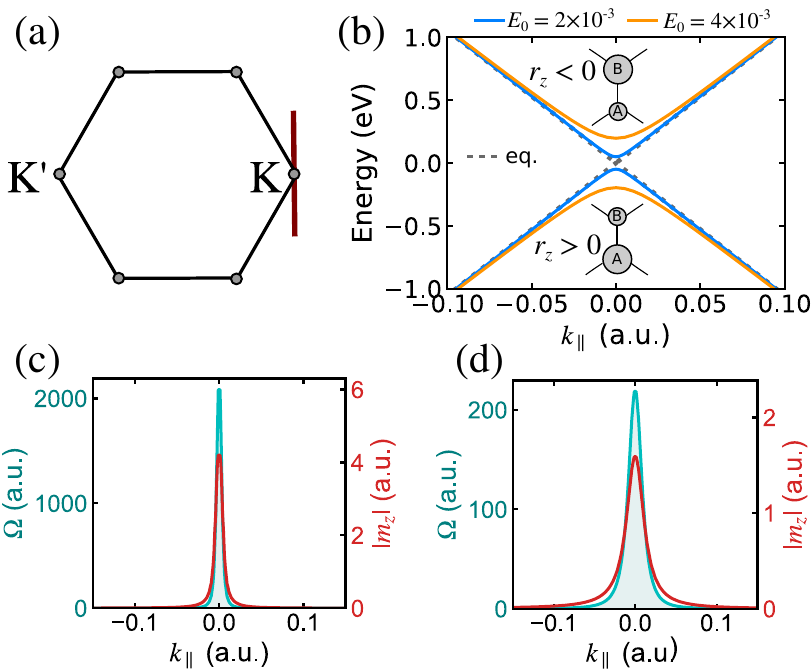}
  \caption{(a) Path in the first Brillouin zone (BZ) close to the K point considered here. (b) Band structure of the effective Hamiltonian $\hat{h}_\mathrm{eff}(\vec{k})$ compared to the equilbrium bands along the path shown in (a). The insets illustrate the  orbital pseudospin. (c),(d): Berry curvature $\Omega(\vec{k})$ and orbital polarization $m_z(\vec{k})$ of the effective lower band (calculated within second-order Brillouin-Wigner theory) for $E_0=2\times 10^{-3}$ (c) and (d) $E_0=4\times 10^{-3}$.
 \label{fig:bands}}
\end{figure}

To connect the photoemission theory to the light-induced topological properties, let us start by discussing the nature of the Floquet state. 
The energy spectrum is obtained from the Floquet Hamiltonian
\begin{align}
\label{eq:floq_ham}
[\hat{\mathcal{H}}_{n n^\prime}]_{j j^\prime}(\vec{k}) &= \frac{1}{T_\mathrm{p}} \int^{T_\mathrm{p}}_0\!d t\, h_{jj^\prime}(\vec{k} - \vec{A}_\mathrm{p}(t)) e^{i (n-n^\prime) \omega_\mathrm{p} t} \nonumber \\ &\quad -
n \omega_\mathrm{p} \delta_{n n^\prime} \delta_{j j^\prime} \ ,
\end{align} 
which captures all steady-state effects including photo-dressing and side bands. A simple physical picture is obtained by applying Brillouin-Wigner theory~\cite{mikami_brillouin-wigner_2016}, which yields the effective Hamiltonian $\hat{h}_\mathrm{eff}(\vec{k})=\sum_{n\ne 0} \hat{\mathcal{H}}_{0 n}(\vec{k}) \hat{\mathcal{H}}_{n 0}(\vec{k})/n \omega_\mathrm{p} $ dressed by virtual absorption and emission processes. Within this picture, a circularly polarized pump field induces next-nearest neighbor hoppings with a complex phase, thus opening a gap $\Delta$ at the two inequivalent Dirac points K, K$^\prime$. The finite gap $\Delta$ corresponds to an occupation imbalance of the two equivalent sublattice sites $j=\mathrm{A,B}$, resulting in a \emph{finite} Berry curvature~\cite{kitagawa_transport_2011}.
Depending on the frequency and field strength of the periodic drive, the effective Hamiltonian can be tailored to be a Chern insulator, with the Chern number determined by the specifics of the pump~\cite{mikami_brillouin-wigner_2016}. For a strong enough high-frequency pump, a nonequilibrium Hall response close to the quantized value can be realized within this model~\cite{mikami_brillouin-wigner_2016,dehghani_out--equilibrium_2015,dehghani_optical_2015}.

The emergence of the gap $\Delta$ is directly connected to the pseudospin properties with respect to the A, B sublattice sites. The effective Hamiltonian can be expressed as $\hat{h}_\mathrm{eff}(\vec{k}) = \vec{D}(\vec{k})\cdot\hat{\gvec{\sigma}}$ ($\hat{\gvec{\sigma}}$ is the vector of Pauli matrices acting on the sublattice space); the vector $\vec{D}(\vec{k})$ characterizes the pseudospin structure of the Hamiltonian.
Expanding around the Dirac points one finds $\vec{D}(\mathrm{K} + \vec{k}) \approx (-v_G k_y, v_G k_x, v^2_G a_0 E_0^2 / \omega^3_\mathrm{p})$ and $\vec{D}(\mathrm{K}^\prime + \vec{k}) \approx (v_G k_y, v_G k_x, -v^2_G a_0 E_0^2 / \omega^3_\mathrm{p})$ with $v_G = 3 J a_0/2$ (lattice constant $a_0$)~\cite{kitagawa_transport_2011}. The gap thus scales as $\Delta \propto D_z(\mathrm{K}) \propto E^2_0/\omega^3_\mathrm{p}$. 
Similarly, the quantum state of the lower (or upper) effective band is characterized by the orbital pseudospin vector $\vec{r}(\vec{k})$; in particular, $r_z(\vec{k})=P_\mathrm{A}(\vec{k})-P_\mathrm{B}(\vec{k})$ measures the occupation difference between the A and B sublattice at a specific point in momentum space. 
The pseudospin $r_z(\vec{k})$ is directly related to the topological properties. 
One can show that the bands are topologically trivial if $D_z(\vec{k})$ does not change sign across the BZ. 
In this case, one finds $r_z(\vec{k})  > 0$ ($r_z(\vec{k})  < 0$) for  $D_z(\vec{k}) > 0$ ($D_z(\vec{k}) < 0$) in the whole BZ, corresponding to a charge-density-wave pattern. In contrast, a topological phase transition is accompanied by $D_z(\vec{k})$ changing sign, resulting in a band inversion. Instead, for graphene in equilibrium and for the idealized limit of vanishing spin-orbit coupling $r_z(\vec{k})=0$, which implies vanishing Berry curvature.
%

Figure~\ref{fig:bands}(b) shows the band structure of the effective Hamiltonian $\hat{h}_\mathrm{eff}(\vec{k})$ for $E_0=2\times 10^{-3}$ and $E_0=4\times 10^{-3}$ a.u. along with the pseudospin properties. The lower (upper) effective band with energy $\en_l(\vec{k})$
($\en_u(\vec{k})$) is characterized by $r_z(\vec{k}) > 0$ ($r_z(\vec{k}) < 0$). In Fig.~\ref{fig:bands}(c),(d) we show the Berry curvature $\Omega(\vec{k})$ of the lower band and the associated \emph{orbital} polarization
\begin{align}
  \label{eq:orbmag}
  m_z(\vec{k}) = -\frac{e}{\hbar}(\en_u(\vec{k}) - \en_l(\vec{k})) \Omega(\vec{k}) 
\end{align} 
from the modern theory of polarization~\cite{souza_dichroic_2008,xiao_berry_2010}. As sketched in Fig.~\ref{fig:setup}(b), this orbital magnetic moment possesses the same symmetry properties as the Berry curvature; in particular, it has the same sign at K and K$^\prime$. In the regime of weak pump driving strength considered here, the pseudospin and topological properties are fully characterized by their behavior in the vicinity of the Dirac points. We remark that stronger fields can induce more complex pseudospin textures~\cite{sentef_theory_2015}, while Floquet sidebands become important.
 
\subsection{Orbital polarization and circular dichroism}

In the absence of magnetic atoms, the induced orbital polarization~\eqref{eq:orbmag} is an intrinsic topological property, which is due to the self-rotation of the underlying Bloch states. This can be understood intuitively by constructing wave packets $|W_{\vec{k}\alpha}\rangle$ from Bloch states in a particular band $\alpha$. The finite spread in real space allows to define the angular momentum $\langle\hat{L}_z \rangle_{\vec{k}\alpha} = \langle W_{\vec{k}\alpha} | \hat{L}_z | W_{\vec{k}\alpha}\rangle$. For a narrow distribution in momentum space, $\langle\hat{L}_z \rangle_{\vec{k}\alpha}$ becomes independent of the specific shape of the wave packet and thus defines the orbital angular momentum of the Bloch states itself~\cite{xiao_berry_2010}, which is connected with the general orbital polarization $m_z(\vec{k}) = e/m \langle\hat{L}_z \rangle_{\vec{k}\alpha}$. 
%

A nonzero orbital magnetic moment $m_z(\vec{k})$ determines the selection rules for photoexcitation properties~\cite{cao_unifying_2018} and thus results in \emph{intrinsic} circular dichroism. In general, circular dichroism arises from different contributions, such as scattering of the photoelectron from the lattice. This \emph{extrinsic} effect is, for instance, responsible for the characteristic dichroic signal from graphene~\cite{gierz_illuminating_2011}. Averaging around high-symmetry points has been suggested as an efficient way of separating the contributions~\cite{schuler_local_2020}. The wave-packet picture provides a direct link to intrinsic circular dichroism in photoemission~\cite{schuler_local_2020}, revealing that the angular momentum $\langle\hat{L}_z \rangle_{\vec{k}\alpha}$ determines the selection rules; vanishing angular momentum corresponds to vanishing dichroism. 

The magnetic moment $m_z(\vec{k})$ is intimately connected to the Berry curvature, as both quantities possess the same symmetry properties~\cite{cho_experimental_2018}. In particular, in the case of two (effective) bands, orbital magnetic moment takes the form of Eq.~\eqref{eq:orbmag} and thus becomes proportional to the Berry curvature. Hence, circular dichroism establishes a link to momentum-resolved topological properties. For two-orbital honeycomb lattice systems like graphene, this link can be found explictly in terms of the orbital pseudospin texture. The leading contribution to the intrinsic dichroism in ARPES becomes~\cite{schuler_local_2020}
\begin{align}
    \label{eq:dichro_pseudo}
  I_\mathrm{CD}(\vec{k}, \en_f) \propto \frac{2}{k} r_z(\vec{k}) k_x a_{\mathrm{CC}} \widetilde{\varphi}_{z}(k,p_\perp) \frac{d}{dk}\widetilde{\varphi}_z(k,p_\perp) \ .
\end{align}
The only missing proportionality factor is the energy conservation. The distance between the two carbon atoms is denoted by $a_\mathrm{CC}$, while $\widetilde{\varphi}_z(k,p_\perp)$ stands for the Fourier-transformed atomic $p_z$ orbital.
Hence, the dichroic signal is directly determined by the pseudospin properties. Within the quasi-static picture of periodically driven graphene outlined above, the induced pseudospin and topological properties are expected to manifest themselves in circular dichroism in trARPES.

\section{Results}

\subsection{Floquet features and scattering processes\label{subsec:transient}}

The simple physical picture based on the effective Brillouin-Wigner Hamiltonian provides a simple description of the opening of the effective bands, but fails to take dynamical processes into account. To establish the link to trARPES under experimentally relevant conditions, we now employ the full time-dependent treatment of the many-body Hamiltonian~\eqref{eq:ham0_td}, with emphasis on the stability of the Floquet physics under e--e and e--ph scattering. Both mechanism give rise to population restribution, dephasing and, in case of e--ph coupling, dissipation. 

\begin{figure}[t]
    \centering
    \includegraphics{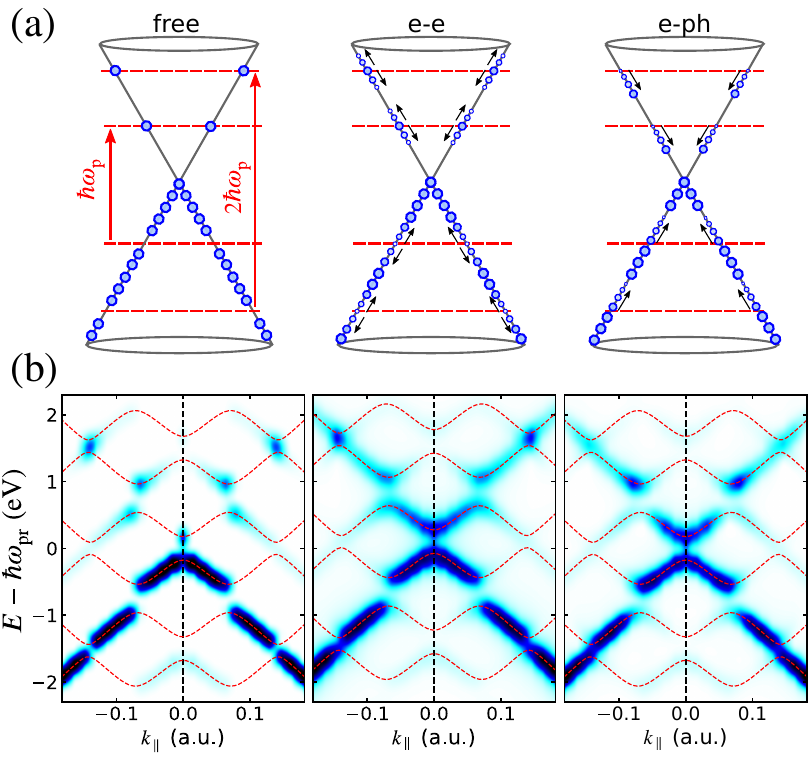}
    \caption{(a) Illustration of the interplay between photoexciation and scattering processes close to a Dirac point. (b) trARPES spectra calculated with the simplified formula~\eqref{eq:tdarpes_simple} with aligned pump ($E_0=4\times 10^{-3}$ a.u.) and probe pulse, for the free system (left), including only e--e (middle) and only e--ph scattering (right panel), respectively.
    The spectra are calculated along the path shown in Fig.~\ref{fig:bands}(a). The energy $E=\en_f - \mu$ is the kinetic energy of the photoelectrons shifted by the chemical potential $\mu=-4.6$~eV.
    The red-dashed lines represent the band structure of the Floquet Hamiltonian~\eqref{eq:floq_ham}. 
    }
    \label{fig:scattering}
\end{figure}

\paragraph*{Redistribution and heating.---}
The dynamics of photoexcitation processes -- especially far from equilibrium -- is strongly influenced by scattering processes. Turning the e--e and e--ph interactions off, vertical transitions induced by the pump pulse lead to isolated points of nonzero population in the upper band, determined by energy and momentum conservation (Fig.~\ref{fig:scattering}(a), left panel). The absorption of energy is limited by these restrictions; resonant driving will induce Rabi oscillations and even decrease the number of excited electrons. The picture changes dramatically when e--e scattering is included (Fig.~\ref{fig:scattering}(a), middle panel), since this leads to a thermalization of electrons (holes) in the upper (lower) band. The balance between the pumping strength and the scattering rate govern the effective Floquet thermalization~\cite{peronaci_resonant_2018,murakami_nonequilibrium_2017}. In the limit of an infinitely long pulse, the system reaches a quasi-thermal distribution with infinite temperature. Similarly, e--ph scattering (Fig.~\ref{fig:scattering}(a), right panel) leads to a redistribution of the excited electrons, thus providing a pathway for further absorption. In contrast to e--e scattering, the dissipative character of e--ph scattering (if the phonons are considered as a heat bath with quasi-infinite heat capacity) give rise to a Floquet steady state~\cite{murakami_nonequilibrium_2017}. 

\begin{figure*}[t]
	\includegraphics[width=0.9\textwidth]{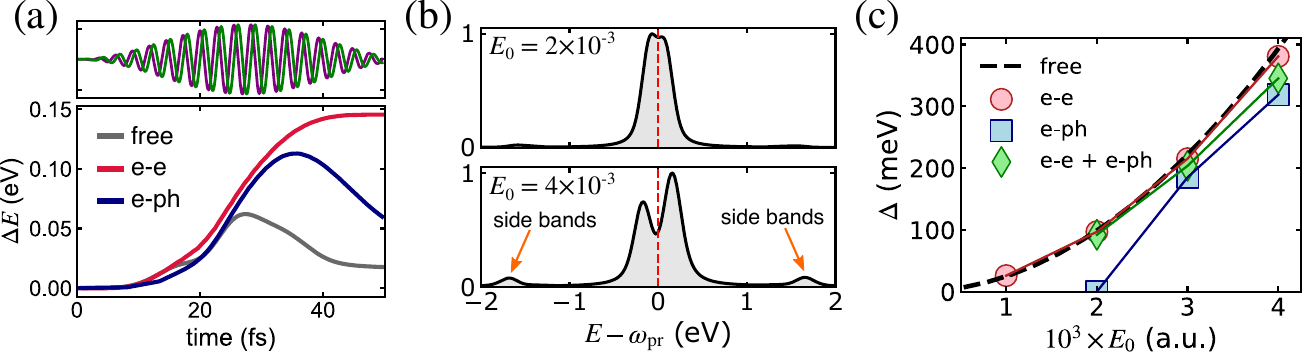}
    \caption{(a) Time-dependent kinetic energy and shape of the pump pulse. 
    The components of the circular pump fields are reported in purple (x) and green (y).
    (b) Photoemission spectra $I(\vec{k}=\mathrm{K},\en_f,\Delta t)$ (cf. Eq.~\eqref{eq:tdarpes_simple}) at the Dirac point, including e--e and e--ph interactions. We use a pump pulse with $N_c=30$ optical cycles.
    (c) Floquet gap $\Delta$ as function of pump field strength, extracted from the spectra in (b) and Fig.~\ref{fig:scattering}(b). 
    For the case of e--ph scattering, the broad spectra do not allow for an unambigous determination of a gap.
 	}
    \label{fig:ekin_gap}
\end{figure*}

For a quantitative picture we computed the time-dependent Green's function as described in Section~\ref{sec:methods}. To exclude that matrix-element effects mask the dynamics discussed here, we simplify the trARPES formula~\eqref{eq:tdarpes} to
\begin{align}
  \label{eq:tdarpes_simple}
  I(\vec{k},\en_f,\Delta t) = &\mathrm{Im}\sum_{j} \int^\infty_0\!d t\! \int^{t}_0\!d t^\prime\, s(t)s(t^\prime)  G^<_{jj}(\vec{k};t,t^\prime) \nonumber \\ &\quad  \times e^{-\iu (\omega_\mathrm{pr} - \en_f)(t-t^\prime)}
\end{align}
for this discussion. We assume overlapping pump and probe pulses ($\Delta t=0$) and express the energy in terms of the binding energy $\en_b$.
The resulting spectra are presented in Fig.~\ref{fig:scattering}(b). The physical picture of scattering processes above is directly applicable to the trARPES spectra. Without any interaction, only electrons at certain momenta are promoted to the excited-state manifold. The band structure captured by Eq.~\eqref{eq:tdarpes_simple} now exhibits Floquet sideband features, described by the Floquet Hamiltonian~\eqref{eq:floq_ham}.
The corresponding band structure is shown by the red-dashed lines in Fig.~\ref{fig:scattering}.
As Fig.~\ref{fig:scattering}(b) shows, the excited-state population is restricted to the avoided crossings of the Floquet bands.

Another important feature of the noninteracting treatment is the peak occupation directly at K just above $\en_b=0$. This is  a manifestion of a "topological hole" in quantum quenches~\cite{schuler_tracing_2017}: driving a topological phase transition and opening the gap, the orbital character is preserved. This effect is confined to the region close to K, where the time evolution is nonadiabatic no matter how slowly the pump pulse is switched on. 


In contrast, the simulation with e--e scattering (Fig.~\ref{fig:scattering}(b), middle panel) yields a considerable redistribution of the occupation. In particular, the population close to K in the effective upper band becomes very pronounced, which is in stark contrast to the noninteracting case. Similar effects are observed as a result of e--ph scattering (Fig.~\ref{fig:scattering}(b), right panel), albeit high-energy features like the peak of the occupation at $\en_b=1.5$~eV are suppressed due to the dissipation. We also note that the spectra of the interacting system align very well with the Floquet bands of the noninteracting system, which indicates that renormalization effects play a minor role (apart from a small energy shift in the presence of e--e interactions). 

As also inferred from Fig.~\ref{fig:scattering}(b), the number of excited electrons is significantly larger if scattering channels are available, giving rise to considerably larger energy absorption. This becomes clear when inspecting the change of kinetic energy per particle $\Delta E$ (Fig.~\ref{fig:ekin_gap}(a)). While the pump pulse injects energy into the free system, this energy is mostly emitted back when the pump envelope approaches zero. This is in stark contrast to the result which includes e--e scattering, which leads to continuous heating and an order of magnitude larger absorption. E--ph scattering has a similar effect, even though the lack of full thermalization and cooling by emitting phonons reduces the kinetic energy.

\paragraph*{Decoherence.---} The pronounced heating and the resulting dephasing effects -- especially for resonant pumping as in graphene -- typically hamper the coherence required for Floquet features~\cite{kandelaki_many-body_2018,sato_floquet_2020}. Fig.~\ref{fig:ekin_gap}(b) shows representative photoemission spectra calculated from Eq.~\eqref{eq:tdarpes_simple}. Note that the broadening of the spectra is mostly due to decoherence effects, as the energy spectrum of the probe pulse is much narrower. 

To investigate how the scattering effects influence the opening of a gap at K (or K$^\prime$), we analyzed $I(\vec{k}=\mathrm{K},\en_b,\Delta t=0)$ by a two-peak Gaussian fit to extract the Floquet gap $\Delta$, presented in Fig.~\ref{fig:ekin_gap}(b). Comparing to the gap $\Delta$ predicted by the noninteracting Brillouin Wigner theory (see Section~\ref{subsec:heff_pseudo}), we find that e--e scattering reduces $\Delta$ only very weakly. As in Ref.~\cite{kandelaki_many-body_2018}, e--e interactions renormalize the band structure in the vicinity of the Dirac points, which increases $\Delta$. However, this effect is compensated by the decoherence due to e--e scattering.
In contrast, e--ph scattering has a strong effect, suppressing the Floquet gap almost completely for $E_0 \le 2\times 10^{-3}$. Consistent with Ref.~\cite{kandelaki_many-body_2018}, increasing the pump field strength stabilizes $\Delta$. To trace the origin of this pronounced effect, we switched off large-momentum e--ph scattering. The resulting spectra are considerably sharper, and the Floquet gap is much more pronounced. This shows that inter-valley scattering is the predominant source of decoherence.

Including both e--e and e--ph scattering, 
this stabilization allows to determine $\Delta$ for $E_0 \ge 2\times 10^{-3}$. The Floquet gap is slightly larger than for e--ph coupling only. These results show that phonons are the major source of decoherence in this regime, while e--e scattering predominantly thermalizes the system. 

Similar to the gap $\Delta$, the Floquet side bands remain stable in the presence of scattering effects, but they are broadened (see Fig.~\ref{fig:ekin_gap}(b)). The stability of the Floquet features (at least for stronger driving) is consistent with the ultrafast time scale of the pump pulse. The period of a single cycle is $2.7$~fs, which is much shorter than any typical scattering time. Decoherence builds up over several pump cycles.

Scattering processes and the resulting heating and decoherence effects also strongly impact transport properties like the Hall response. In particular, decoherence was identified as key factor~\cite{sato_floquet_2020} to understand ultrafast transport experiments~\cite{mciver_light-induced_2020}, albeit on an empirical level. Investigating ultrafast scattering processes as captured by our theory thus provides a microscopic perspective on transport properties.

\subsection{Ultrafast Hall response\label{subsec:hall}}

The light-induced topological state in the considered regime is described by the effective Floquet Hamiltonian $\hat{h}_\mathrm{eff}(\vec{k})$ (see Section~\ref{subsec:heff_pseudo}), which yields a Chern number of $C=1$ for the lower effective band. In the quasi-static picture, the system should thus exhibit a quantized Hall response in the limit of low effective temperatures~\cite{mikami_brillouin-wigner_2016}. However, the nonequilibrium situation in a pump-probe setup renders a straightforward detection and interpretation of the time-dependent Hall current difficult. Decoherence due to scattering processes will reduce the Floquet gap (see Section~\ref{subsec:transient}) and suppress the Hall response~\cite{sato_microscopic_2019-1,sato_light-induced_2019-1}. Increasing the pump strength stabilizes the Floquet features at the cost of stronger pump-induced heating.

\begin{figure}[b]
\centering
\includegraphics[width=0.8\columnwidth]{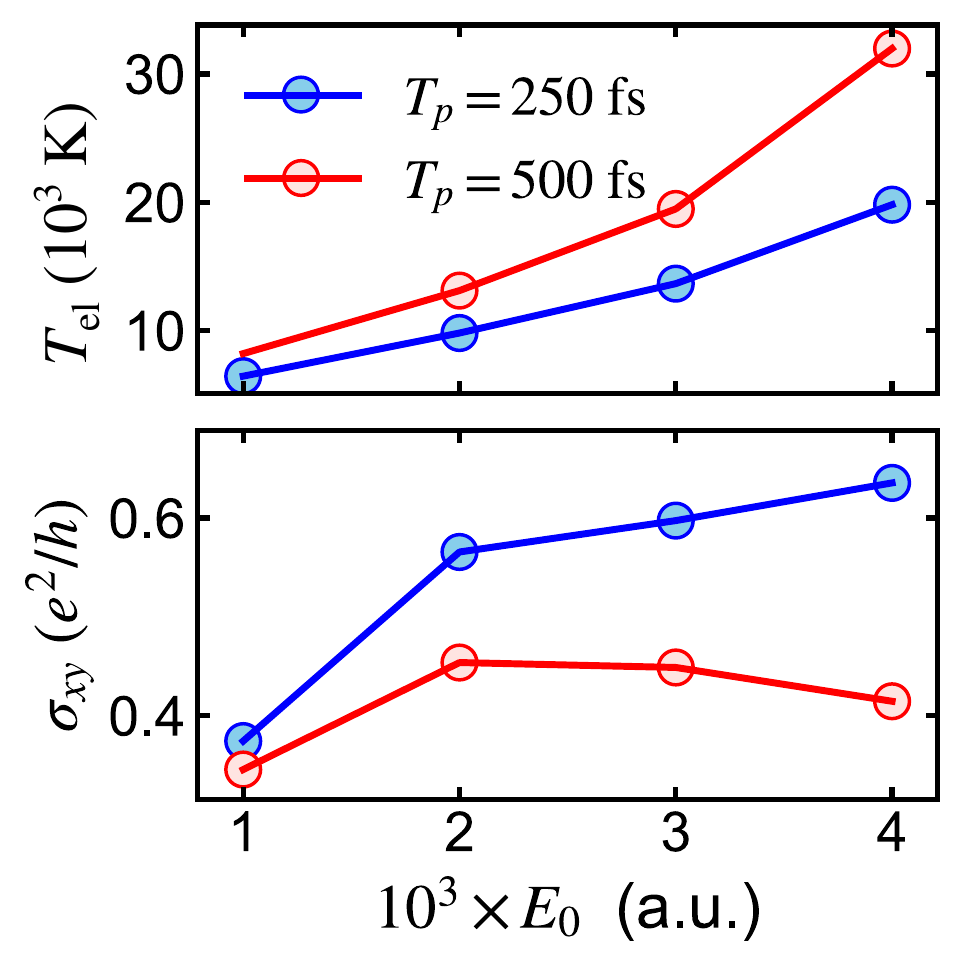}
\caption{Upper panel: Electronic temperature $T_\mathrm{el}$ obtained from the single-temperature model~\eqref{eq:tel_single}
for pump pulse duration $T_p=250$~fs (blue) and $T_p=500$~fs (red) as a function of the pump field strength $E_0$. Lower panel: corresponding Hall response. \label{fig:hall} }
\end{figure}

The anomalous Hall response under pumping and including e--e and e--ph scattering can, in principle, be obtained from the GKBA time propagation. However, state-of-the-art experimental techniques enable the detection of ultrafast transient currents on the picosecond time scale, which is still a relatively long time scale for microscopic many-body simulations, so that a direct comparison is difficult.

\paragraph*{Steady-state model.---}
As explained in Section~\ref{subsec:transient}, the e--e and e--ph interactions primarily lead to a redistribution of the occupation, while the effective band structure is governed by the free Floquet Hamiltonian. Moreover, the distribution is quasi-thermal with respect to the Floquet bands. In this situation, the Floquet nonequilibrium steady-state (NESS) formalism~\cite{aoki_nonequilibrium_2014} provides an excellent description. Details are presented in Appendix~\ref{sec:floquetarpes}. In essence, we assume that each lattice site of graphene is coupled to a thermalizing bath (coupling strength $\gamma$), which is characterized by an effective temperature $T_\mathrm{eff}$. This setup corresponds to a metallic substrate; however, here we treat it as a generic pathway for dissipation and dephasing.

The balance between absorption and dissipation determines the occupation of the Floquet bands. The NESS formalism yields the Green's function $G^<_{jj^\prime}(\vec{k},t,t^\prime)$ (which is now periodic in both time arguments).
Inserting this expression into Eq.~\eqref{eq:tdarpes_simple}, and assuming a infinitely long pump and probe pulses, yields
\begin{align}
    \label{eq:floqarpes_simple}
    I(\vec{k},\en_f) \propto \mathrm{Im} \sum_{j}  G^<_{jj}(\omega_\mathrm{pr}- \en_f)\ ,
\end{align}
where 
\begin{align}
    \label{eq:gles_freq}
    G^<_{jj^\prime}(\omega) &= \frac{1}{T_\mathrm{p}} \int^{T_\mathrm{p}}_0\! d t_\mathrm{av}
    \int^\infty_{-\infty}\! d t_\mathrm{rel}\, e^{i \omega t_\mathrm{rel}} \nonumber \\ &\quad\quad\times G^<_{jj^\prime}\left(t_\mathrm{av} + \frac{t_\mathrm{rel}}{2}, t_\mathrm{av} - \frac{t_\mathrm{rel}}{2}\right) \ .
\end{align}
The NESS spectra obtained from Eq.~\eqref{eq:floqarpes_simple} can be considered a very good fitting model for the trARPES spectra; a quantitative comparison is shown in Section~\ref{subsec:robust}. Following Ref.~\cite{mikami_brillouin-wigner_2016}, the Hall response can be directly obtained from the NESS model (see Appendix~\ref{sec:floquetarpes} for details). Fixing $\gamma$ to match the line width of the trARPES spectra in Fig.~\ref{fig:ekin_gap}(b), the effective temperature $T_\mathrm{eff}$ remains the only free parameter.

\paragraph*{Thermalization and effective temperature.---}
To connect to the time-dependent microscopic treatment (including thermalization due to the scattering) and access the picosecond time scale, we employ a single-temperature model for the electronic temperature $T_\mathrm{el}(t)$ adopted from Ref.~\cite{caruso_photoemission_2020}:
\begin{align}
	\label{eq:tel_single}
	\frac{d}{dt} T_\mathrm{el}(t) = \frac{\mathcal{I}(t)}{\alpha c_\mathrm{el}(T_\mathrm{el}(t))} \ ,
\end{align}
where $c_\mathrm{el}(T)$ denotes the electronic heat capacity at temperature $T$, while $\mathcal{I}(t)$ represents the envelope of the intensity of the pump pulse. The parameter $\alpha$ is adjustable. Note that we ignore the phonon contribution here, as the rapid thermalization of the phonon subsystem prevents cooling of electrons during the pump pulse. The model obtained by solving Eq.~\eqref{eq:tel_single} is then fitted to the electronic temperature obtained from the GKBA simulation for varying length of the pump pulse. 

With this model at hand, we can extrapolate $T_\mathrm{el}(t)$ to longer time scales. Figure~\ref{fig:hall} shows the electronic temperature for a pulse duration of $T_p=250$~fs and $T_p=500$~fs. The system heats up considerably; one finds a scaling $T_\mathrm{el}\sim E^2_0$ and roughly $T_\mathrm{el}\sim T_p$ in the considered regime. The Hall response $\sigma_{xy}$ (defined as the time average of the Hall current) is shown in the lower panel in Fig.~\ref{fig:hall}. For $T_p=250$~fs, $\sigma_{xy}$ increases monotonically with $E_0$, although a saturation sets in for $E_0 \le 2\times 10^{-3}$. The maximum value reached is $\sigma_{xy}\approx 0.63 e^2/h$, which corresponds to $\sim 32 \%$ of the quantized value $\sigma^{(0)}_{xy}=2 e^2/h$. 
Increasing the pump duration to $T_p=500$~fs, the heating effects dominate for larger $E_0$, thus suppressing the Hall response with growing $E_0$. The quantized value $\sigma^{(0)}_{xy}$ cannot be approached in this regime. 

In general, the nonequilibrium Hall response contains a topological contribution due to the induced Berry curvature (see Section~\ref{subsec:heff_pseudo}) and a contribution arising from a probe-induced population imbalance~\cite{sato_microscopic_2019-1}. The latter has been found to dominate for weak pump field strengths, as used here~\cite{sato_microscopic_2019-1}. Note that the NESS formalism employed here includes both contributions at the level of linear response. Isolating the Berry curvature contribution is a considerable task; this is where circular dichroism and the energy resolution of trARPES can provide valuable complementary insights.

\begin{figure*}[t]
    \centering
    \includegraphics{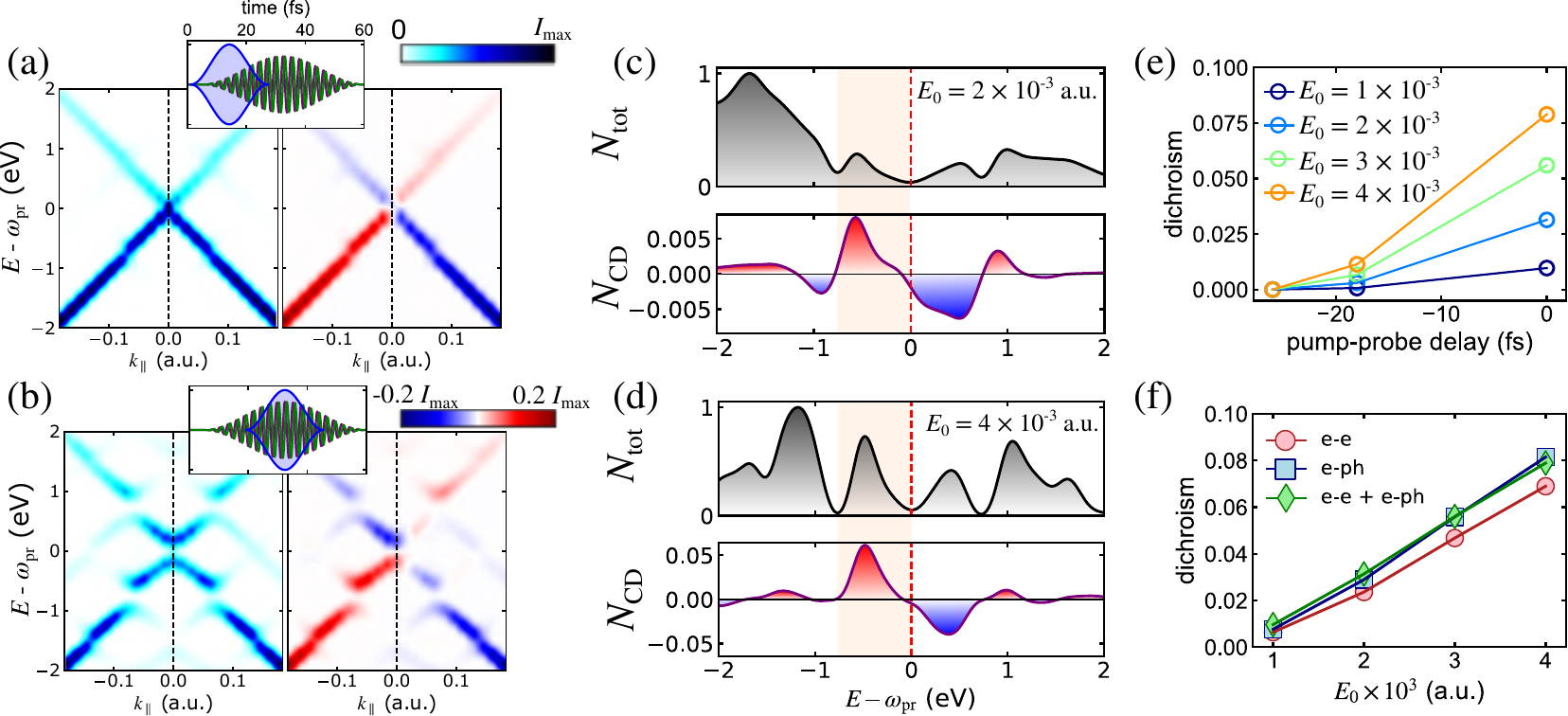}
    \caption{(a), (b): Build-up of Floquet features and circular dichroism in trARPES for $\Delta t = -18$~fs (a) and $\Delta t=0$ (b), calculated along the path depicted in Fig.~\ref{fig:bands}(a). All spectra have been obtained including e--e and e--ph scattering. The energy $E=\en_f-\mu$ is measured with respect to the chemical potential $\mu$.
    (c), (d): Momentum-integrated unpolarized and dichroic trARPES spectra (over a disk around K with $k_r=0.25$~a.u. radius) for $E_0=2\times 10^{-3}$ (c) and $E_0=4\times 10^{-3}$ (d). The shaded background indicates the range of energy integration for (e) and (f). (e) Relative integrated dichroic signal as a function of the pump-probe delay. (f) Dichroic signal as a function of the pump field strength.  }
    \label{fig:hubb_phon_cdad}
\end{figure*}

\subsection{Time-resolved photoemission and circular dichroism\label{subsec:tdarpes}}

Now we investigate how the induced Berry curvature and pseudospin texture manifest themselves in the circular dichroism. To this end, we employ the full trARPES expression~\eqref{eq:tdarpes} including photoemission matrix elements and laser dressing of the final states. The time-dependent Green's function entering Eq.~\eqref{eq:tdarpes} is computed taking both e--e scattering and e--ph scattering into account. 

Figure~\ref{fig:hubb_phon_cdad}(a)--(b) shows the build-up of the photo-dressed band structure, captured by the unpolarized intensity $I_\mathrm{tot}(\vec{k},E,\Delta t)$ (the energy $E$ is  the kinetic energy of the photoelectron shifted by the chemical potential $\mu$), and the corresponding dichroic signal $I_\mathrm{CD}(\vec{k},E,\Delta t)$. In the initial phase of the pump-induced dynamics ($\Delta t <0$), Floquet features like a gap opening or sidebands are hardly visible, although a kink at $E-\omega_\mathrm{pr}=-0.75$~eV indicates the onset of transient photodressing (Fig.~\ref{fig:hubb_phon_cdad}(a)). The portion of the pump pulse overlapping with the probe pulse (see inset in Fig.~\ref{fig:hubb_phon_cdad}(a)) is broad in frequency space; therefore, electrons are excited nonresonantly and redistributed by e--e and e--ph scattering. The dichroic signal resembles the equilibrium case~\cite{liu_visualizing_2011,gierz_graphene_2012}, but with positive dichroism ($I_\mathrm{CD}(\vec{k},E,\Delta t) > 0$) for $E-\omega_\mathrm{pr} < 0$. For aligned pump and probe pulses (Fig.~\ref{fig:hubb_phon_cdad}(b)), the unpolarized spectrum combines features of the cases of e--e and e--ph scattering in Fig.~\ref{fig:scattering}(b). By switching off the corresponding phase factor in Eq.~\eqref{eq:tdarpes}, we find that LAPE effects have little influence, apart from a slight enhancement of the sideband intensity relative to the zero-photon effective band. Inspecting $I_\mathrm{CD}(\vec{k},E,\Delta t=0)$, a clear asymmetry becomes apparent. In particular, below $E-\omega_\mathrm{pr}=0$ close to the K point, positive dichroism (i.\,e. photoemission by a LCP probe pulse) dominates.

To investigate the dichroism in more detail, we integrate the trARPES signal over a small region in momentum space in the vicinity of K,
\begin{align}
    \label{eq:Ntotcd}
    N_\mathrm{tot/CD}(E-\omega_\mathrm{pr}) = \int\! d\vec{k}\, I_\mathrm{tot/CD}(\vec{k},\en_f,\Delta t=0) \ ,
\end{align}
presented in Fig.~\ref{fig:hubb_phon_cdad}(c)--(d) for $E_0=2\times 10^{-3}$ and $E_0=4\times 10^{-3}$. Close to $E-\omega_\mathrm{pr}=0$, the dichroism is pronounced and positive (negative) below (above) the Fermi energy. This sign change corresponds directly to the behavior of the pseudospin in Fig.~\ref{fig:bands}(b). It is remarkable that our full one-step theory -- which includes scattering of the photoelectron from the lattice -- is in line with the simple physical picture outlined in Section~\ref{subsec:heff_pseudo}. While intricate final-state effects have a profound impact on the concrete angular distribution of the dichroism, quantities integrated around high-symmetry points are more sensitive to \emph{intrinsic} circular dichroism related to topological properties. We have confirmed this picture for several systems in Ref.~\cite{schuler_local_2020}. Therefore, the dichroism observed in Fig.~\ref{fig:hubb_phon_cdad}(c)--(d) arises due to the orbital magnetization and follows the proportionality to the pseudospin~\eqref{eq:dichro_pseudo}.

To quantify the circular dichroism relative to the unpolarized signal, we integrated $P_\mathrm{tot/CD} = \int\!d\en N_\mathrm{tot/CD}(\en)$ over the orange-shaded range in Fig.~\ref{fig:hubb_phon_cdad}(c)--(d). The ratio $P_\mathrm{CD}/P_\mathrm{tot}$ is shown in Fig.~\ref{fig:hubb_phon_cdad}(e) as a function of the pump-probe delay, which confirms that the build-up of the dichroism follows the pump envelope. Inspecting $P_\mathrm{CD}/P_\mathrm{tot}$ for $\Delta t =0$ (Fig.~\ref{fig:hubb_phon_cdad}(f)), we find a roughly linear dependence on the pump field $E_0$. Although the pseudospin behaves as $r_z(\mathrm{K})\propto E_0^2$ (which can be seen from Brillouin-Wigner theory), the heating effects, which increase with $E_0$, result in an overall linear behavior. As the system absorbs more energy if only e--e scattering is present, the dichroism is slightly reduced as compared to the system with only e--ph scattering. Apart from such subtle effects, the dichroism is remarkably stable against dephasing and scattering. Note that the dichroism is sizable for $E_0=2\times 10^{-3}$, where almost no Floquet gap can be observed (see Fig.~\ref{fig:ekin_gap}(b)).

\subsection{Robustness of the circular dichroism\label{subsec:robust}}

The stability of the circular dichroism against interaction effects -- in contrast to the Floquet gap -- is a striking feature. To corroborate that this conclusion is not limited to the simplified model of e--e correlations, or an artefact of the specific treatment in this work, we have performed calculations within the NESS formalism (see Appendix~\ref{sec:floquetarpes} for details). Inserting the Green's function obtained from Eq.~\eqref{eq:gles_freq} into Eq.~\eqref{eq:tdarpes} and assuming an infinitely long pump and probe pulse yields
\begin{align}
    \label{eq:floqarpes}
    I(\vec{k},\en_f) \propto \mathrm{Im} \sum_{jj^\prime} M^*_j(\vec{k},p_\perp) G^<_{jj^\prime}(\omega_\mathrm{pr}- \en_f) M_{j^\prime}(\vec{k},p_\perp) \ .
\end{align}
Note that we have neglected LAPE effects here. The Green's function is fully determined by the properties of the bath, characterized by the coupling strength $\gamma$ and the effective temperature $T_\mathrm{eff}$. 

\begin{figure}[b]
    \centering
    \includegraphics{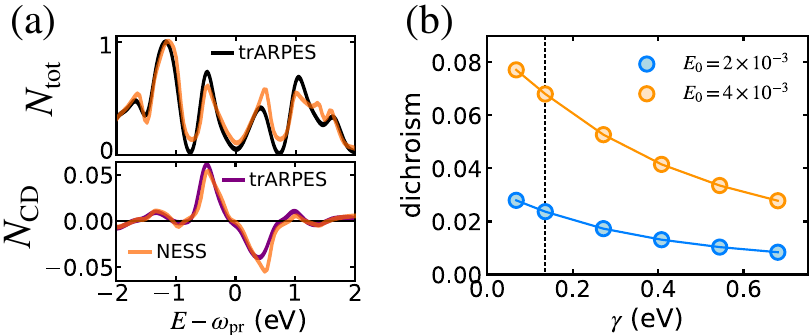}
    \caption{(a) Comparison between the momentum-integrated unpolarized (upper) and dichroic (lower panel) signal for $E_0=4\times 10^{-3}$ a.u. obtained from trARPES (same as in Fig.~\ref{fig:hubb_phon_cdad}(d)) and from the nonequilibrium steady-state (NESS) formalism. The effective temperature is set to $T_\mathrm{eff} = 1/30$~a.u.. (b) Energy-integrated dichroism (extracted as in Fig.~\ref{fig:hubb_phon_cdad}(f)) as a function of the bath coupling strength $\gamma$. The dashed vertical line indicates the value of $\gamma$ in (a). }
    \label{fig:ness_vs_trarpes}
\end{figure}

Figure~\ref{fig:ness_vs_trarpes}(a) illustrates that the Floquet NESS description of ARPES~\eqref{eq:floqarpes} provides a very good approximation of the full trARPES treatment~\eqref{eq:tdarpes} for appropriate parameters $\gamma$ and $T_\mathrm{eff}$. Even though the agreement for the momentum-integrated signal $N_\mathrm{tot}$ is good, deviations indicate that the system exhibits a nonthermal distribution for overlapping pump and probe pulses
\footnote{This also holds when including e--e interactions only.}.
The dichroic signal $N_\mathrm{CD}$, however, agrees very well. On this basis, we can now increase the bath coupling strength $\gamma$ -- which also sets the dephasing time scale -- and investigate the robustness of the circular dichroism. Performing an analysis as for Fig.~\ref{fig:hubb_phon_cdad}(f), we calculate the relative energy-integrated dichroic signal for increasing $\gamma$. The result is presented in Fig.~\ref{fig:ness_vs_trarpes}(b). The dichroism stays robust over a large range of dissipation strength; more than doubling $\gamma$ compared to the realistic value used in Fig.~\ref{fig:ness_vs_trarpes}(a) roughly reduces the dichroism by a factor of two. We remark that $\gamma$ also sets the linewidth of the photoemission spectra and thus captures the decoherene effects discussed in Section~\ref{subsec:transient}.

We also calculate photoemission spectra according to Eq.~\eqref{eq:floqarpes} for the values of $\gamma$ from Fig.~\ref{fig:ness_vs_trarpes}(b). The Floquet gap gets strongly suppressed for increasing dephasing $\gamma$, consistent with Ref.~\cite{sato_light-induced_2019-1}, up to a point where no gap can be observed anymore. Nevertheless, as Fig.~\ref{fig:ness_vs_trarpes}(b) demonstrates, the circular dichroism stays robust even in this strongly dissipative regime.

\section{Summary and discussion}

We have presented a detailed investigation of the topological properties of graphene pumped with circularly polarized light under realistic conditions. Within the simple picture based on an effective renormalized Hamiltonian, a gap opens at the Dirac points, thus giving rise to nonzero Berry curvature and orbital polarization. In the considered regime of $\hbar\omega_\mathrm{p}=1.5$~eV pump photon energy and for realistic field strength, graphene becomes a Floquet-Chern insulator. 

Finding definite manifestations of the induced topological state in experiments has been a challenge. The opening of a gap -- as first reported in trARPES experiments on Bi$_2$Se$_3$~\cite{wang_observation_2013,mahmood_selective_2016} -- would be a clear signature of the effective Floquet bands. Based on our time-dependent atomistic calculations, including e--e and e--ph interactions, we showed that the Floquet bands are formed, but broadened by the dephasing due to the interaction. Both e--e and e--ph scattering are essential to capture the pronounced occupation of the excited bands. Heating effects would be severly underestimated by simulations which lack these scattering channels. 
The Floquet gap is found to be relatively stable against interaction effects for larger pump field strengths, although e--ph coupling -- predominantly inter-valley scattering -- gives rise to significant dephasing. For weak to moderate field strengths ($E_0 \le 2\times 10^{-3}$ a.u.), the Floquet gap is hardly visible, which implies that the opening of a gap is not a useful criterion for experiments in the considered regime. 

Besides the Floquet gap, the anomalous Hall current measured during the pump pulse -- similar to the experiment in Ref.~\cite{mciver_light-induced_2020} -- can provide insights into the induced toplogical state. Extrapolating to feasible time scales by an effective temperature model we find very pronounced heating effects. These heating effects (which increase with the field strength) compete with the stabilization of the Floquet gap, thus suppressing the Hall response far below the quantized value for realistically long pump pulses. In particular, for a pulse duration of $T_p=500$~fs, we find that the heating effects dominate for stronger pump fields. Furthermore, the Hall response also contains a contribution arising from a population imbalance, which is difficult to discern from the contribution originating from the induced Berry curvature.

This is where trARPES can provide valuable insights. The energy resolution allows to observe the effective bands and their occupation even in a hot state. Morever, measuring circular dichroism provides a direct link to the topological state. \emph{Intrinsic} dichroism arises due to the orbital magnetization~\cite{schuler_local_2020}, which is proportional to the Berry curvature in the simple effective model. For graphene, in particular, the dichroism is a direct map of the induced pseudospin texture, which is intimately connected to the topological state. Combining accurate one-step calculations of the photoemission matrix elements with atomistic time-dependent simulations provides a state-of-the-art approach to trARPES and the circular dichroism in particular. We have shown that the thus obtained band-resolved dichroism is in line with the pseudospin properties and is sizable even when the Floquet gap cannot be observed. Furthermore, the dichroism is robust against scattering effects and dissipation, which is corroborated by a steady-state model. 

Measuring circular dichroism in ARPES -- accompanied by a predictive theory -- thus provides a tool for tracing topological properties in and out of equilibrium in an unprecedented way. In particular, the build-up of light-induced states~\cite{claassen_all-optical_2016,hubener_creating_2017,topp_all-optical_2018} and their topological character can be traced with full band (and even spatial~\cite{wong_pulling_2018,cucchi_microfocus_2019}) resolution in real time, which will boost the discovery and understanding of transient topological phenomena. 

\section*{Acknowledgments}
We acknowledge helpful discussion with Shunsuke A. Sato. 
M.\,S. and T.\,P.\,D. acknowledge financial support from the U.\,S. Department of Energy (DOE), Office of Basic Energy Sciences, Division of Materials Sciences and Engineering, under contract no. DE-AC02-76SF00515. Furthermore, this work was supported by the
Swiss National Science Foundation via NCCR MARVEL
and the European Research Council via ERC-2015-AdG-694097 and ERC Consolidator Grant No.~724103. The Flatiron Institute is a division of the Simons Foundation. 
M.~S. thanks the Alexander von Humboldt
Foundation for its support with a Feodor Lynen scholarship and the Department. 
M.~A.~S.~acknowledges financial support by the DFG through the Emmy Noether program (SE 2558/2-1). 

\appendix

\section{Tight-binding Hamiltonian and orbitals\label{sec:tbdetails}}

We describe the electronic structure of graphene at the level of the next-nearest-neighbor
tight-binding (TB) model, defined by
\begin{align}
	\label{eq:ham0}
  \hat{H}_0 = \sum_{\vec{k}}\sum_{j,j^\prime,\sigma} h_{j j^\prime}(\vec{k}) \hat{c}^\dagger_{\vec{k}j \sigma} \hat{c}_{\vec{k} j^\prime \sigma} \ .
\end{align}
Here, $\hat{c}^\dagger_{\vec{k} j \sigma}$ ($\hat{c}_{\vec{k} j \sigma}$) creates (annihilates) an electron with momentum $\vec{k}$ and spin $\sigma$; $j$ labels the sublattice site within the unit cell. Employing a compact matrix notation, the Hamiltonian is constructed in the TB approximation as
\begin{align}
  \label{eq:ham0_matrix}
  \vec{h}(\vec{k}) = \begin{pmatrix}  0 & g(\vec{k}) \\ g^*(\vec{k}) & 0 \end{pmatrix}
\end{align}
with
\begin{align}
g(\vec{k}) = -J e^{\iu \vec{k}\cdot \gvec{\tau}}\left(1 + e^{-\iu \vec{k}\cdot \vec{a}_2} + e^{-\iu \vec{k}\cdot (\vec{a}_1+\vec{a}_2)}\right) \ ,
\end{align}
where $\vec{a}_{1,2}$ denote the lattice vectors and $\gvec{\tau}=\vec{t}_\mathrm{B}-\vec{t}_\mathrm{A}$ the vector connecting the sublattice sites. The hopping amplitude is chosen as $J=2.628$~eV.

The Bloch states $\psi_{\vec{k}\alpha}(\vec{r})$ are obtained by the Wannier representation
\begin{align}
  \label{eq:wannrep}
  \psi_{\vec{k}\alpha}(\vec{r}) &=  \frac{1}{\sqrt{N}}\sum_{\vec{R}}
  \sum_j C_{\alpha j}(\vec{k}) e^{\iu\vec{k}\cdot (\vec{R}+\vec{t}_j)} 
  w_j(\vec{r}-\vec{R}) \nonumber \\
  &\equiv \sum_j C_{\alpha j}(\vec{k})\phi_{\vec{k}j}(\vec{r}) \ ,
\end{align}
where the coefficients $C_{\alpha j}(\vec{k})$ are the eigenvectors of the Hamiltonian~\eqref{eq:ham0_matrix}. The Wannier orbitals are approximated by Gaussian wave-functions of the type
\begin{align}
  w_{j}(\vec{r}) = C_j z e^{-\alpha_j
  (\vec{r}-\vec{t}_j)^2} \ .
\end{align}
The parameters $C_j$ and $\alpha_j$ are fitted to atomic orbitals. 

\section{One-step calculation of matrix elements\label{sec: one-step}}

The photoemission intensity is governed by Fermi's Golden rule, given by
\begin{align}
  \label{eq:fermigolden}
  I(\vec{p},\en_f) \propto \left|\langle \chi_{\vec{p},p_\perp} |
  \hat{\epsilon}\cdot \hat{D} | \psi_{\vec{k}\alpha}\rangle\right|^2
  \delta(\en_{\vec{k}\alpha} + \hbar \omega - \en_f) \ .
\end{align}
Here, the photon energy is given by $\hbar
\omega$, and $\en_f=(\vec{p}^2+p^2_\perp)/2$ is the energy of the 
photoelectron final state $|\chi_{\vec{p},p_\perp}\rangle$. The matrix
element of the dipole operator $\hat{D}$ and the
polarization direction $\hat{\epsilon}$ determine the selection
rules. The in-plane momentum $\vec{p}$ is identical to the
quasi-momentum $\vec{k}$ up to a reciprocal lattice vector. While formally equivalent, the choice of the gauge for the transition operator $\hat{D}$ plays an important role in developing accurate approximations. In this work, we use the momentum operator $\hat{D} = \hat{\vec{p}} = (\hbar/\iu) \gvec{\nabla}$. However, for capturing effects such as circular dichroism, accurate final states $|\chi_{\vec{p},p_\perp}\rangle$ are required. For instance, approximating the final states by plane waves, the circular dichroism vanishes. Therefore, we compute $|\chi_{\vec{p},p_\perp}\rangle$ explictly as eigenstates of a model potential. In particular, we construct a muffin-type scattering potential of the form
\begin{align}
  \label{eq:muffintin}
  V(\vec{r}) = \sum_{\vec{R}} v_0(|\vec{r}-\vec{R}|) \ ,
\end{align}
where the sum runs over all lattice sites. The spherical atom-centered potential is modelled by a smoothed box-like dependence $v_0(r) = -V_0/(1 + \exp[a_0(r-r_0)])$. The parameters $V_0$, $a_0$ and $r_0$ are adjusted to approximate the \emph{ab initio} photoemission spectra (see below). 

The final states are Bloch states with respect to the in-plane momentum, while they obey time-reversed LEED asymmptotic boundary conditions in the out-of-plane ($z$) direction. Thus, it is convenient to expand the final states as
\begin{align}
	\label{eq:ileed_expand}
  \chi_{\vec{p},p_\perp}(\vec{r}) = \sum_{\vec{G}}
  e^{\iu(\vec{p}+\vec{G})\cdot \vec{r}} \xi_{\vec{p},p_\perp;\vec{G}}(z) \ .
\end{align}
The photoelectron momentum $\vec{p}$ is identical to the crystal momentum $\vec{k}+\vec{G_0}$ of the initial Bloch states due to in-plane momentum conservation (up to a reciprocal lattice vector $\vec{G}_0$). Assuming photoemision from the first BZ ($\vec{G}_0=0$),
the expansion coefficients in Eq.~\eqref{eq:ileed_expand} are fixed by 
\begin{align}
	\label{eq:ileed_bound}
	\xi_{\vec{k},p_\perp;\vec{G}}(z) &\rightarrow e^{\iu \vec{k}\cdot\vec{r}} + \sum_{\vec{G}} R_{\vec{G}} e^{-\iu (\vec{k}+\vec{G})\cdot\vec{r}} \ , \quad (z\rightarrow \infty) \nonumber \\
	\xi_{\vec{k},p_\perp;\vec{G}}(z) &\rightarrow \sum_{\vec{G}} T_{\vec{G}} \rightarrow e^{\iu (\vec{k} + \vec{G})\cdot\vec{r}}\ , \quad (z\rightarrow -\infty)   \ , 
\end{align}
where $R_{\vec{G}}$ and $T_{\vec{G}}$ are reflection and transmission coefficients, respectively. Expanding the potential in plane waves,
\begin{align}
	V(\vec{r}) = \sum_{\vec{G}} e^{\iu \vec{G}\cdot \vec{r}} V_{\vec G}(z) \ ,
\end{align}
the final states~\eqref{eq:ileed_expand} are determined by the Schr\"odinger equation
\begin{align}
  \label{eq:scattwf}
  \sum_{\vec{G}^\prime} &\left[\left(-\frac{\partial^2_z}{2} + 
\frac{(\vec{p}+\vec{G})^2}{2} \right)\delta_{\vec{G},\vec{G}^\prime} +
  V_{\vec{G}-\vec{G}^\prime}(z)\right] \xi_{\vec{p},p_\perp;\vec{G}^\prime}(z) \nonumber
  \\ &=
  \left(\frac{\vec{p}^2}{2}+\frac{p^2_\perp}{2}\right) \xi_{\vec{p},p_\perp;\vec{G}}(z) \ .
\end{align}
We solve Eq.~\eqref{eq:scattwf} together with the boundary condition~\eqref{eq:ileed_bound} employing the renormalized Numerov method as in Ref.~\cite{schuler_nuclear-wave-packet_2014}.

After obtaining the final states, the matrix elements $M_\alpha(\vec{k},p_\perp) = \langle \chi_{\vec{k},p_\perp}  | \hat{\epsilon}\cdot\hat{D}|\psi_{\vec{k}\alpha} \rangle$ are computed by
\begin{align}
  \label{eq:mepe2}
  M_\alpha(\vec{k},p_\perp) = \sum_{\vec{G}} \hat{\epsilon}\cdot
  (\vec{k}+\vec{G})  \int^\infty_{-\infty}\dd z\,
  \xi^*_{\vec{p},p_\perp;\vec{G}}(z)
  \phi_{\vec{k}\alpha;\vec{G}}(z) \ ,
\end{align}
where $\phi_{\vec{k}\alpha;\vec{G}}(z)$ denote the plane-wave expansion coefficients of the Bloch states~\eqref{eq:wannrep}. 

\begin{figure}[t]
  \centering
  \includegraphics[width=\columnwidth]{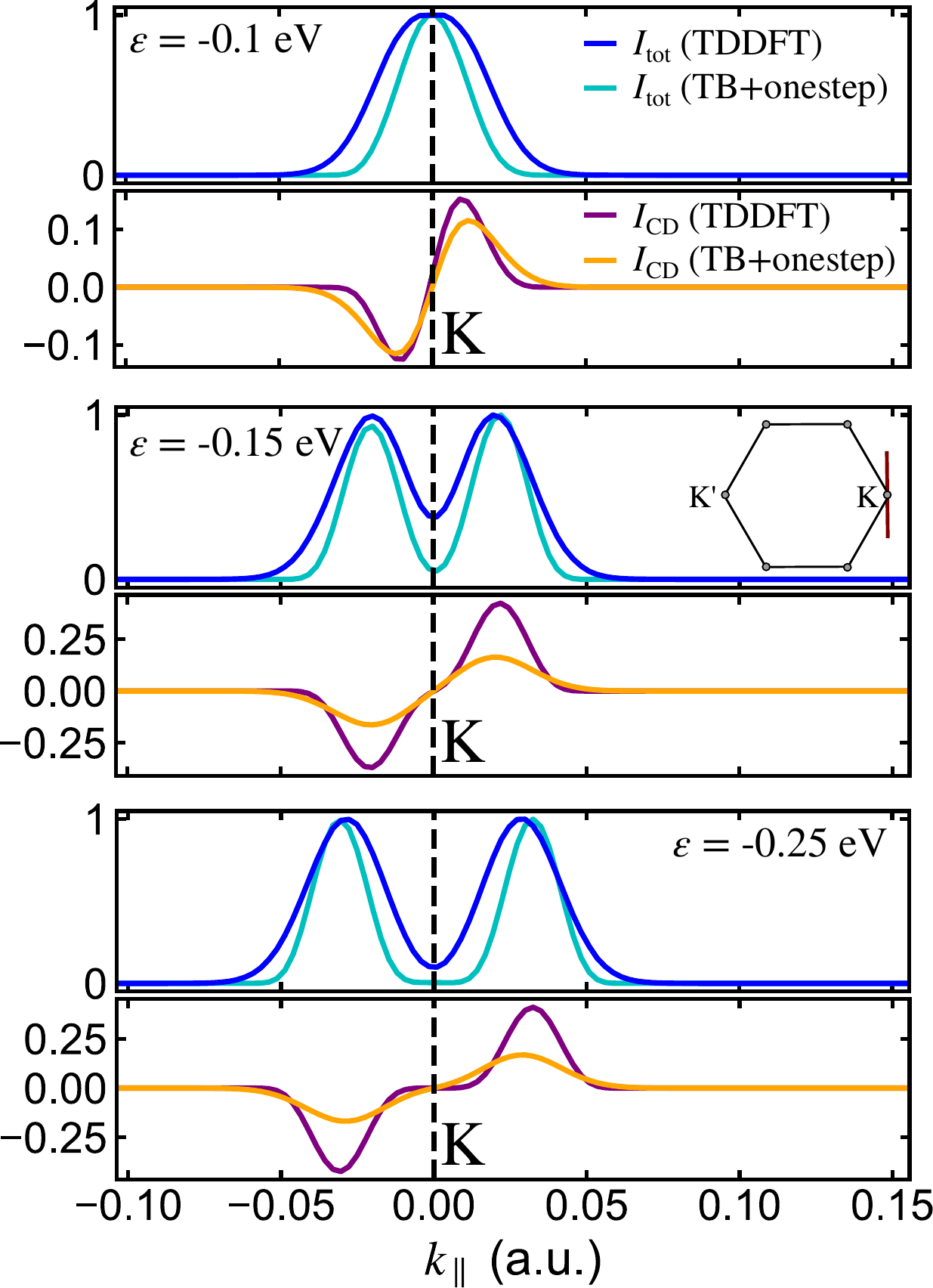}
  \caption{Comparison of the ARPES intensity $I(\vec{k}, \en_f)$ for different binding energies within the \emph{ab initio} method and the TB+ one-step approach. We show the characteristic path in the BZ orthogonal to $\Gamma$-K passing through K (see inset). \label{fig:benchmark}}
\end{figure}

The thus calculated matrix elements are benchmarked against \emph{ab initio} calculations based on TDDFT (analogous to Ref.~\cite{schuler_local_2020}. We find the best agreement of the resulting ARPES spectra with the first-principle results for $V_0=3.0$, $a_0=5$ and $r_0=1$ (atomic units). As a characteristic benchmark, we computed the total intensity $I_\mathrm{tot}(\vec{k},\en_f)$ and the circular dichroism $I_\mathrm{CD}(\vec{k},\en_f)$ along the path 
shown in the inset in Fig.~\ref{fig:benchmark}. As is known from theory~\cite{schuler_local_2020} and experiment~\cite{gierz_graphene_2012,liu_visualizing_2011}, this is where the circular dichroim is most pronounced (while it vanishes in the $\Gamma$-K direction). 

Fig.~\ref{fig:benchmark} shows $I_\mathrm{tot}(\vec{k},\en_f)$ and $I_\mathrm{CD}(\vec{k},\en_f)$ within the TB+ one-step theory and compares it to the first-principle calculations. Except for the exact magnitude of the circular dichroism, the TB+ one-step approach matches the TDDFT results very well, thus endorsing it as excellent method for qualitative behavior (especially close to the Dirac point).

\section{Time-dependent nonequilibrium Green's functions calculations \label{sec:numdetails}}

We treat the dynamics in pumped graphene including e--e scattering as well as e--ph coupling within the framework of the td-NEGF approach, based on the the single-particle Green's function (GF) on the Kadanoff-Baym contour $\mathcal{C}$:
\begin{align}
	\label{eq:contourgf}
	G_{j j^\prime,\sigma}(\vec{k}; t,t^\prime) = -\iu \langle T_{\mathcal{C}} \hat{c}_{\vec{k}j\sigma}(t)
	\hat{c}^\dagger_{\vec{k}j^\prime\sigma}(t^\prime)\rangle \ .
\end{align} 
Here, where $T_{\mathcal{C}}$ denotes the contour ordering symbol. Since the spin-orbit coupling is negligibly small in graphene, we drop the spin index in what follows. The contour GF~\eqref{eq:contourgf} obeys the equation of motion
\begin{align}
	\label{eq:contoureom}
	\left(\iu \partial_t - \vec{h}^\mathrm{MF}(\vec{k},t)\right) \vec{G}(\vec{k}; t,t^\prime)
	  &= \delta_{\mathcal{C}}(t,t^\prime) \nonumber \\
	  &\quad + \int_{\mathcal{C}}\!\dd \bar{t}\, \gvec{\Sigma}(\vec{k}; t,\bar{t}) 
	  \vec{G}(\vec{k}; \bar{t},t^\prime)  \ .
\end{align}
Here, we have employed the compact matrix notation. The self-energy $\gvec{\Sigma}(\vec{k}; t,t^\prime)$ captures all interaction effects beyond the mean-field (MF) Hamiltonian $\vec{h}^\mathrm{MF}(\vec{k},t)$. Projecting onto observables times using the Langreth rules transforms the  equation of motion~\eqref{eq:contoureom} into the usual Kadanoff-Baym equations (KBEs). Solving the KBEs poses a considerable computational challenge, as the computational effort grows as $N^3_t$ with $N_t$ time steps. To reduce the numerical effort and the memory demands, we employ the generalized Kadanoff-Baym ansatz (GKBA). The GKBA tranforms the two-time KBEs to the single-time kinetic equation for the single-particle density matrix $\gvec{\rho}(\vec{k},t)$:
\begin{align}
\label{eq:KBE_transport}
\partial_t \gvec{\rho}(\vec{k},t) + i [\vec{h}^\mathrm{MF}(\vec{k},t),\gvec{\rho}(\vec{k},t)] = - (\vec{I}(\vec{k},t)+\mathrm{h.\,c.}) \ ,
\end{align}
where the collision integral $\vec{I}(\vec{k},t)$ is defined by
\begin{align}
\label{eq:collint}
\vec{I}(\vec{k},t) &= \int^t_{-\infty}\! d\bar{t} \big( \gvec{\Sigma}^<(\vec{k};t,\bar{t}) \vec{G}^\mathrm{A}(\vec{k};\bar{t},t) \nonumber \\ &\quad + \gvec{\Sigma}^\mathrm{R}(\vec{k};t,\bar{t}) 
\vec{G}^<(\vec{k};\bar{t},t) \big) .
\end{align}
Correlations of the initial state $\gvec{\rho}(\vec{k};t=0)$ are built in by adiabatic switching: at $t=-\infty$, the equilibrium density matrix is determined by the MF treatment, while correlation effects are gradually incorporated by replacing $\gvec{\Sigma}(\vec{k};t,t^\prime)\rightarrow f(t)f(t^\prime)\gvec{\Sigma}(\vec{k};t,t^\prime)$ with a smooth switch-on function $f(t)$.
However, Eqs.~\eqref{eq:KBE_transport} and \eqref{eq:collint} are not closed in terms of $\gvec{\rho}(\vec{k},t)$ since, in principle, information on the whole two-time dependence of the GF enters the collision integral \eqref{eq:collint}.
Within the GKBA, the two-time dependence, which captures spectral information, is approximated by
\begin{subequations}
\label{eq:gkba_lesgtr}
\begin{align}
\vec{G}^<(\vec{k};t,t^\prime) &= - \vec{G}^\mathrm{R}(\vec{k};t,t^\prime)\gvec{\rho}(\vec{k},t^\prime) + \gvec{\rho}(\vec{k},t) \vec{G}^\mathrm{A}(\vec{k};t,t^\prime), \\
\vec{G}^>(\vec{k};t,t^\prime)  &= \vec{G}^\mathrm{R}(\vec{k};t,t^\prime) \bar{\gvec{\rho}}(\vec{k},t^\prime) - \bar{\gvec{\rho}}(\vec{k},t) \vec{G}^\mathrm{A}(\vec{k};t,t^\prime) \ , 
\end{align}
\end{subequations}
where $\bar{\gvec{\rho}}(\vec{k},t) = 1 - \gvec{\rho}(\vec{k},t^\prime)$.
Here we approximate $\vec{G}^\mathrm{R}(\vec{k};t,t^\prime)$ by the MF GF
\begin{align}
\label{eq:eom_gret}
\left(i\partial_t - \vec{h}^\mathrm{MF}(\vec{k},t)\right) \vec{G}^\mathrm{R}(\vec{k};t,t^\prime) = \delta(t-t^\prime) \ .
\end{align}

\subsection{Electron-electron interaction\label{subsec:eeint}}

As has been shown in Ref.~\cite{schuler_optimal_2013}, graphene can be treated in good approximation as effective Hubbard model with $U\approx 1.6 |J|$, which we adopt in this work. Thus, we consider
\begin{align}
\label{eq:ham_ee_hubb}
	\hat{H}_{\mathrm{e-e}}  = \frac{U}{2}\sum_{\vec{R}}\sum_{j,\sigma} \left(\hat{n}_{\vec{R}j \sigma}-\frac12\right)
	\left(\hat{n}_{\vec{R}j \bar\sigma}-\frac12\right) \ ,
\end{align}
where $\hat{n}_{\vec{R}j\sigma}$ is the density operator for unit cell $\vec{R}$.

The value for $U$ is clearly in the weakly interacting regime. Therefore, we employ the second-order expansion in the Coulomb interaction (second-Born approximation, 2BA) for the self-energy:
\begin{align}
\label{eq:sigma_ee_hubb}
	\Sigma^{\mathrm{e-e},\gtrless}_{j j^\prime}(\vec{k}; t,t^\prime) = \frac{U^2}{N^2_k} &\sum_{\vec{q},\vec{p}} 
	G^\gtrless_{j j^\prime}(\vec{k}-\vec{q}; t,t^\prime) G^\gtrless_{j j^\prime}(\vec{q}+\vec{p}; t,t^\prime)
	\nonumber \\ &\quad \times
	G^\lessgtr_{j^\prime j}(\vec{p}; t^\prime,t) \ .
\end{align}
Here, $N_k$ denotes the number of points sampling the BZ.

\subsection{Electron-phonon coupling\label{subsec:epc}}

We also include e-ph interactions, which are modelled by the Hamiltonian
\begin{align}
\label{eq:ham_eph}
	\hat{H}_{\mathrm{e-ph}} = \frac{1}{\sqrt{N_k}}\sum_{\vec{q},\nu} \frac{1}{\sqrt{M_C\omega_{\vec{q}\nu}}} \sum_{\vec{k}} \sum_{j l,\sigma} \Gamma^\nu_{j l}(\vec{q}) \hat{c}^\dagger_{\vec{k}-\vec{q} j\sigma} \hat{c}_{\vec{k} l \sigma} \hat{X}_{\vec{q}\nu} \ ,
\end{align}
where we include the phonon modes $\nu\in \{\mathrm{LA, TA, LO, TO}\}$. $\omega_{\vec{q}\nu}$ stands for their dispersion; $M_C$ is the mass of the carbon atom. The phonon coordinate operator is defined by $\hat{X}_{\vec{q}\nu} = (\hat{b}_{\vec{q}\nu} + \hat{b}^\dagger_{-\vec{q}\nu})/\sqrt{2}$.

Systematic studies and transport experiments~\cite{sohier_phonon-limited_2014,park_electronphonon_2014} have demonstrated the feasiblity of weak-coupling treatment. Hence, we employ the (non-selfconsistent) Midgal approximation. The e-ph contribution to the self-energy is then given by
\begin{align}
	\Sigma^{\mathrm{e-ph},\gtrless}_{j j^\prime}(\vec{k}; t,t^\prime) = \frac{\iu}{N_k} \sum_{\vec{q},\nu} \frac{1}{M_C \omega_{\vec{q}\nu}} &\sum_{l l^\prime}  \Gamma^\nu_{j l}(\vec{q}) G^\gtrless_{l l^\prime}(\vec{k}-\vec{q}; t,t^\prime) \nonumber \\ &\quad  \Gamma^\nu_{l^\prime j^\prime}(\vec{q}) D^\gtrless_\nu(\vec{q}; t, t^\prime) \ .
\end{align}
Here, $D^\gtrless_\nu(\vec{q}; t, t^\prime)$ denotes the free phonon GF.

 The e-ph coupling matrix elements $\Gamma^\nu_{j l}(\vec{q})$ are computed from the symmetry of the phonon modes and the Bloch states. In this work, we adopt the canonical modes from~\cite{piscanec_kohn_2004}, while the e-ph couplings are taken from the TB model from Ref.~\cite{sohier_phonon-limited_2014}. For completeness, we gather the for the couplings below:
\begin{subequations}
\begin{align}
	\gvec{\Gamma}^\mathrm{TA}(\vec{q}) = |\vec{q}|\begin{pmatrix} 2 \alpha & \beta_\mathrm{A} \vec{e}_{-}(\vec{q})^2 \\
	\beta_\mathrm{A} \vec{e}_{+}(\vec{q})^2 & 2 \alpha
	 \end{pmatrix} \ ,
\end{align}
\begin{align}
	 \gvec{\Gamma}^\mathrm{LA}(\vec{q}) = |\vec{q}|\begin{pmatrix} 0 & \beta_\mathrm{A} \vec{e}_{-}(\vec{q})^2 \\
	\beta_\mathrm{A} \vec{e}_{+}(\vec{q})^2 &0
	 \end{pmatrix} \ ,
\end{align}
\begin{align}
	 \gvec{\Gamma}^\mathrm{LO}(\vec{q}) = \iu \begin{pmatrix} 0 & \beta_\mathrm{O} \vec{e}_{+}(\vec{q}) \\
	-\beta_\mathrm{O} \vec{e}_{-}(\vec{q}) & 0 
	 \end{pmatrix}\ ,
\end{align}
\begin{align}
	 \gvec{\Gamma}^\mathrm{TO}(\vec{q}) = \begin{pmatrix} 0 & -\beta_\mathrm{O} \vec{e}_{+}(\vec{q}) \\
	-\beta_\mathrm{O} \vec{e}_{-}(\vec{q}) & 0 
	 \end{pmatrix}\ .
\end{align}
\end{subequations}
Here, $\vec{e}_{\pm}(\vec{q}) = (q_x \pm \iu q_y)/|\vec{q}|$. For the constants $\alpha$, $\beta_\mathrm{A}$ and $\beta_\mathrm{O}$ we adopt the $GW$ values from Ref.~\cite{sohier_phonon-limited_2014}.

\subsection{Spectral corrections \label{subsec:spectral_corr}}

The GKBA underestimates self-energy effects for the two-time dependence of the GF. Therefore, we correct the retarded GF by the static correlation correction from Ref.~\cite{latini_charge_2014} by solving 
\begin{align}
    \label{eq:static_corr_gret}
    \left(i\partial_t - \vec{h}^\mathrm{MF}(\vec{k},t) - \widetilde{\gvec{\Sigma}}(\vec{k},t) \right) \widetilde{\vec{G}}^\mathrm{R}(\vec{k}; t,t^\prime) = \delta(t-t^\prime) \ .
\end{align}
Here, the effective one-time self-energy is approximated by 
\begin{align}
    \label{eq:static_corr_sigma}
    \widetilde{\gvec{\Sigma}}(\vec{k},t) = \int\! d \bar{t}\, \gvec{\Sigma}^\mathrm{R}(\vec{k},t,t-\bar{t}) \ .
\end{align}
Tests showed that these corrections have little influence on the dynamics of $\gvec{\rho}(\vec{k},t)$. Therefore, we employ the correction in a "one-shot" fashion: After obtaining $\gvec{\rho}(\vec{k},t)$ for all time steps, we construct the lesser and greater GFs according to Eq.~\eqref{eq:gkba_lesgtr}, substitute them into Eq.~\eqref{eq:static_corr_sigma} and compute the retarded GF from Eq.~\eqref{eq:static_corr_gret}. Finally, the corrected lesser GF is obtained from Eq.~\eqref{eq:gkba_lesgtr}, replacing $\vec{G}^\mathrm{R}\rightarrow \widetilde{\vec{G}}^\mathrm{R}$.

\subsection{Numerical details}

The GKBA calculations were performed with a highly accurate, in-house computer code. All collision integrals~\eqref{eq:collint} are computed using fifth-order Gregory quadrature~\cite{schuler_nessi:_2019}, while the equation of motion~\eqref{eq:KBE_transport} is solved with a fifth-order Adams-Moulton predictor-corrector scheme. We used $N_t=4500$ to $N_t=5600$ equidistant time points and a time step of $h=0.5$~a.u. (convergence has been checked). The full first BZ is sampled by a $N_k=96\times 96$ grid in momentum space.

\section{Floquet steady-state formalism\label{sec:floquetarpes}}

The Floquet nonequilibrium steady-state (NESS) formalism is powerful tool for describing the dynamically equilibrated balance of absorption, scattering and dissipation~\cite{aoki_nonequilibrium_2014}. Here we consider the noninteracting graphene system, where each lattice site is coupled to a fermionic bath, charaterized by an embedding self-energy~\cite{stefanucci_nonequilibrium_2013}. To determine the steady-state, one first solves for retarded Floquet GF
\begin{align}
    \label{eq:floq_gret}
    \left[\hat{\mathcal{G}}^\mathrm{R}(\vec{k},\omega)\right]^{-1} = \omega - \hat{\mathcal{H}}(\vec{k}) - \hat{\Sigma}^\mathrm{R}(\omega)  \ ,
\end{align}
where $\hat{\mathcal{H}}(\vec{k})$ denotes the matrix representation of the Floquet Hamiltonian~\eqref{eq:floq_ham} in the combined space of orbitals and Floquet indices. For the retarded self-energy, we invoke the wide-band limit approximation (WBLA)
$\Sigma^\mathrm{R}_{n j, n^\prime j^\prime}(\omega) = - i \delta_{n n^\prime} \delta_{j j^\prime} \gamma/2$. The parameter $\gamma$ describes the coupling strength. The lesser component of the self-energy representing the occupation of the bath is given by
\begin{align}
    \label{eq:floq_gles}
    \Sigma^<_{n j, n^\prime j^\prime}(\omega) = i \delta_{n n^\prime} \delta_{j j^\prime} \gamma f(\omega - \mu + n \omega_\mathrm{p}) \ ,
\end{align}
where $f(\omega)$ denotes the Fermi distribution with inverse temperature $\beta=1/T_\mathrm{eff}$; $\mu$ is the chemical potential of the reservoir, which is assumed to be aligned with the chemical potential of undoped graphene. The lesser Floquet GF is then determined by the Keldysh equation
\begin{align}
    \label{eq:floq_keldysh}
    \hat{\mathcal{G}}^<(\vec{k},\omega) = \hat{\mathcal{G}}^\mathrm{R}(\vec{k},\omega) \hat{\Sigma}^<(\omega) \left[\hat{\mathcal{G}}^\mathrm{R}(\vec{k},\omega) \right]^\dagger \ .
\end{align}
Finally, the physical GF is obtained by switching to the two-time representation
\begin{align}
    \label{eq:phys_gles}
    G^<_{jj^\prime}(\vec{k};t,t^\prime) &= \sum_{n n^\prime} \int^{\omega_\mathrm{p}/2}_{-\omega_\mathrm{p}/2}\frac{d \omega}{2 \pi}
    \mathcal{G}^<_{nj, n^\prime j^\prime}(\vec{k};\omega) \nonumber \\ &\quad\quad\times e^{-i \omega (t-t^\prime)} e^{-i n\omega_\mathrm{p} t}e^{i n^\prime\omega_\mathrm{p} t^\prime}\ .
\end{align}
The thus obtain GF is substituted into Eq.~\eqref{eq:gles_freq}, which yields the steady-state photoemission expression~\eqref{eq:floqarpes}.

The Hall response is calculated (ignoring vertex corrections) as in Ref.~\cite{mikami_brillouin-wigner_2016}:
\begin{widetext}
\begin{align}
    \label{eq:hall_floq}
    \sigma_{xy}(\omega) = \frac{1}{\omega}\frac{2}{N_k S_c}\mathrm{Re} \mathrm{Tr}\sum_{\vec{k}}
    \int^{\omega_\mathrm{p}/2}_{-\omega_\mathrm{p}/2}\!\frac{d\omega^\prime}{2\pi} 
    \Big(\hat{\mathscr{v}}_x(\vec{k}) \hat{\mathcal{G}}^\mathrm{R}(\vec{k},\omega^\prime+\omega)
    \hat{\mathscr{v}}_y(\vec{k})\hat{\mathcal{G}}^<(\vec{k},\omega^\prime) + \hat{\mathscr{v}}_x(\vec{k}) \hat{\mathcal{G}}^<(\vec{k},\omega^\prime+\omega)
    \hat{\mathscr{v}}_y(\vec{k})\hat{\mathcal{G}}^\mathrm{A}(\vec{k},\omega^\prime)\Big)
\end{align}
\end{widetext}
in the limit $\omega\rightarrow 0$. Here, the velocity matrix elements in Floquet representation are defined by
\begin{align}
    \mathscr{v}_{\alpha, nj,n^\prime j^\prime}(\vec{k}) = \frac{1}{T_\mathrm{p}} \int^{T_\mathrm{p}}_0\! d t\, \frac{\partial}{\partial k_\alpha} h_{j j^\prime}(\vec{k}-\vec{A}_\mathrm{p}(t)) 
        e^{i (n-n^\prime) \omega_\mathrm{p} t}.
\end{align}
At a given coupling strength $\gamma$ (and fixed pump parameters), the only adjustable parameter is the effective temperature $T_\mathrm{eff}$, which can be different from the electronic temperature $T_\mathrm{el}$. To obtain $T_\mathrm{eff}$ corresponding to a certain $T_\mathrm{el}$, we computed the averaged kinetic energy (without a pulse) by 
\begin{align}
    E_\mathrm{kin} = \frac{1}{N_k}\sum_{\vec{k}}\sum_{j j^\prime} h_{j j^\prime}(\vec{k}) \frac{1}{T_\mathrm{p}}\int^{T_\mathrm{p}}_0\! d t\, G^<_{j j ^\prime}(\vec{k},t,t) 
\end{align}
from the physical GF~\eqref{eq:phys_gles} as a function of $T_\mathrm{eff}$. Comparing to the dependence the dependence $E_\mathrm{kin}(T_\mathrm{el})$ in thermal equilibrium then allows for determining the relation between $T_\mathrm{eff}$ and $T_\mathrm{el}$.


%

\end{document}